\documentclass[12pt]{article}   
\usepackage{epsfig}
\newcommand{\mysection}{\setcounter{equation}{0}\section}
\renewcommand{\theequation}{\thesection.\arabic{equation}}
\def\beq{\begin{equation}}
\def\eeq{\end{equation}}
\def\beqa{\begin{eqnarray}}
\def\eeqa{\end{eqnarray}}

\newlength{\dinwidth} \newlength{\dinmargin}
\begin{document}

\begin{center}
{\Large \bf Collinear and soft gluon corrections to Higgs production at NNNLO}
\end{center}
\vspace{2mm}
\begin{center}
{\large Nikolaos Kidonakis}\\
\vspace{2mm}
{\it Kennesaw State University, Physics \#1202\\
1000 Chastain Rd., Kennesaw, GA 30144-5591}
\end{center}

\begin{abstract}
I present analytical expressions for the 
collinear and soft gluon corrections to Higgs production via 
the process $b{\bar b}\rightarrow H$ as well as $gg \rightarrow H$
through next-to-next-to-next-to-leading order (NNNLO). 
The soft corrections are complete while the 
collinear corrections include leading and some subleading  
logarithms. Numerical results at the Tevatron and the LHC 
are presented, primarily for $b{\bar b}\rightarrow H$. 
It is shown that the collinear terms greatly improve the 
soft and virtual approximation at next-to-leading order (NLO) and 
next-to-next-to-leading order (NNLO), 
especially when subleading terms are
included. The NNNLO collinear and soft corrections provide
significant enhancements to the total cross section. 
I also provide expressions for the 
collinear and soft corrections through NNNLO 
for  the related Drell-Yan process. 
\end{abstract}

\thispagestyle{empty} \newpage \setcounter{page}{2}

\mysection{Introduction}

The search for the Higgs boson \cite{Higgs} is one of the most important goals 
at the Tevatron and the LHC colliders \cite{HiggsWG}. The main Standard Model 
production channel at these colliders is $gg \rightarrow H$. However, 
the channel $b{\bar b}\rightarrow H$ can also be important 
in the Minimal Supersymmetric Standard Model (MSSM) and other theories 
beyond the Standard Model.
In fact, in the MSSM $b{\bar b}\rightarrow H$ dominates $gg \rightarrow H$ 
at high $\tan\beta$, where $\tan\beta$ is the ratio of the vacuum expectation 
values for the two Higgs doublets.
Complete results through NNLO are known for 
$gg \rightarrow H$ \cite{HK1,CAKM,RSvN}, 
$b{\bar b}\rightarrow H$ \cite{HKbb} and 
the related Drell-Yan processes $q{\bar q} \rightarrow V$ \cite{DY}.

Hard-scattering cross sections can receive large corrections from 
soft-gluon contributions near threshold \cite{GS,CT,KS}. Such corrections 
have been studied for a wide class of cross sections (for a review 
see \cite{NKuni}) where the soft corrections dominate the cross section. 
For the process $gg \rightarrow H$, however, it was shown in \cite{KLS} 
that the soft (or even the soft and virtual) corrections are inadequate
to serve as a good approximation of the complete cross section at NLO.
Collinear corrections have to be added to obtain a reasonable 
approximation. This was also shown at NNLO in \cite{RSvN}. 
We verify these results for $gg \rightarrow H$, and we investigate 
the soft and collinear approximations for the process  
$b{\bar b}\rightarrow H$. We find that 
at both NLO and NNLO the soft, or soft and virtual, approximation 
is inadequate and that collinear logarithms must be added to 
improve the approximation.
 
The processes $q {\bar q} \rightarrow V$, $b{\bar b}\rightarrow H$, and 
$gg\rightarrow H$, present a unique opportunity to study complete 
soft-gluon corrections even at NNNLO. Although partial NNNLO results 
have appeared for other processes, such as top quark production 
\cite{NKNNNLO,NKsingletop},
the complications involved in such processes are far greater. 
The processes $q {\bar q} \rightarrow V$, $b{\bar b}\rightarrow H$, and 
$gg\rightarrow H$, are much more amenable and easy to study (relative 
to, for example, heavy quark or jet production) for two 
distinct reasons:

(i) The color structure of these processes is trivial. 
At lowest order the processes are 
electroweak and the final state does not involve any colored particles 
(quarks or gluons) so the color vertex is trivial. 
This is to be contrasted with processes such as heavy quark \cite{KS} 
or jet \cite{KOS} production where the complicated color structure requires 
matrices in the study of the soft corrections.

(ii) The kinematics of these processes in the calculation of total cross 
sections are trivial.
Again this is to be contrasted with heavy quark or jet production where
the kinematics is more complicated and where the structure and form of the 
soft-gluon terms depends crucially on the kinematics.

It should thus come as no surprise that Drell-Yan and Higgs production
are more well understood and known than other hard scattering processes 
(save deep inelastic scattering).

A unified approach to calculating the soft-gluon corrections for arbitrary 
hard-scattering processes through NNNLO is available in \cite{NKNNNLO}. 
We note that related processes involving the Higgs to which the 
formalism has been applied are charged Higgs production via 
$bg \rightarrow t H^-$ \cite{NKNNNLO,NKchiggs} and 
the process $bg \rightarrow bH$ \cite{FRJ}.
The NNNLO master formula in Ref. \cite{NKNNNLO} 
is quite long because it is structured so it can address all the possible 
complications in color structure and kinematics of a general process. 
Because of the two reasons mentioned above the application of the 
formula to $q {\bar q} \rightarrow V$, $b{\bar b}\rightarrow H$, and 
$gg\rightarrow H$, results in much simpler expressions. 

For the process $b {\bar b} \rightarrow H$ the complete NNLO corrections 
were presented in \cite{HKbb}; however the analytic results presented there 
did not include the scale-dependent terms, and the color structure of the 
corrections (which is useful to know from a theoretical standpoint and for 
resummation) was not made explicit. In this paper I present the complete 
soft-gluon corrections and the collinear logarithms beyond leading accuracy
at NLO, NNLO, and NNNLO including the full color factors and the 
scale-dependent terms. Analytic expressions for the scale-independent terms
in the soft-gluon corrections through NNNLO were also presented in 
Ref. \cite{VR} including the full color structure. The results in this paper 
are in complete agreement with \cite{VR}.

Studies have been performed on the soft-gluon corrections
to the process $gg \rightarrow H$ as well as the Drell-Yan process 
at NNNLO in \cite{VR,MVNNNLO}. However the analytical results for the 
scale-dependent terms were not presented.  
In this paper we derive the full NNNLO soft corrections  for 
$gg \rightarrow H$ and $q {\bar q} \rightarrow V$, including explicitly
terms with the factorization and renormalization scales, 
thus verifying and extending the results presented in \cite{VR,MVNNNLO}.
Furthermore, we calculate collinear logarithms through NNNLO.

The goals of this paper can be summarized as follows:

(1) For the process $b {\bar b} \rightarrow H$: To extend earlier results 
for the NNLO soft and collinear corrections; to investigate the 
contribution of soft, virtual, and collinear corrections to the total 
cross section and determine how well they approximate the complete 
corrections at NLO and NNLO;
to derive at NNNLO the complete soft-gluon corrections and the 
collinear corrections beyond 
leading logarithm accuracy; and to provide for the first time a detailed study 
through NNNLO of the numerical impact of these corrections at the 
Tevatron and the LHC. 

(2) To provide the complete analytical results of the soft-gluon corrections, 
including scale-dependent terms, and the collinear corrections beyond 
leading accuracy through NNNLO for the process $gg \rightarrow H$ and 
for Drell-Yan production. Since these processes have been studied extensively 
before we do not study numerical applications of the results (except in 
a limited way for $gg \rightarrow H$ for comparison with 
$b {\bar b} \rightarrow H$).

(3) To study the effects of subleading terms in the soft-gluon 
expansion and in collinear corrections and make connections with 
earlier results for various processes. 

The paper is organized as follows. In Section 2 we provide a short overview 
of the formalism for the resummation of soft and collinear logarithms 
in Higgs and Drell-Yan production. In Section 3 we derive explicit 
analytical expressions for the soft and collinear corrections for the 
process $b {\bar b} \rightarrow H$. Detailed numerical results are  
provided in Section 4 for $b {\bar b} \rightarrow H$ at the Tevatron and 
the LHC, together with a few results for $gg \rightarrow H$.
The conclusions are in Section 5. Appendix A collects detailed expressions 
for the numerous quantities needed in the calculations. 
Appendices B and C contain explicit analytical expressions for the soft 
and collinear corrections for the Drell-Yan process and for 
Higgs production via $gg \rightarrow H$, respectively.

\mysection{Resummation of soft and collinear logarithms}

It is well known that soft and collinear logarithms in hard-scattering 
cross sections exponentiate \cite{GS,CT,KS}.
In this section we review the resummation formalism for these corrections.
The resummation of soft and collinear logarithms is carried out 
in moment space.
We define moments of the partonic cross section by
${\hat\sigma}(N)=\int dz \, z^{N-1} {\hat\sigma}(z)$,
with $z=M^2/s$. 
Here $M$ is a hard scale which we will take to be the Higgs mass $m_H$ 
for Higgs production and the invariant mass $\sqrt{Q^2}$ of the produced 
dilepton pair for Drell-Yan production, while 
$s=(p_1+p_2)^2$ with $p_i$ the momenta of the incoming partons.
At partonic threshold, $z=1$.
$N$ is the moment variable and the logarithms of $N$ exponentiate.
The resummed partonic cross section in moment space can then be written as
\cite{NKNNNLO}
\beqa
{\hat{\sigma}}_{res}^S(N) &=&
\exp\left[ -2 \int^1_0 dz \frac{z^{N-1}-1}{1-z}\;
\left \{\int^1_{(1-z)^2} \frac{d\lambda}{\lambda}
A_i\left(\alpha_s(\lambda M^2)\right)
+{\nu}_i \left[\alpha_s((1-z)^2 M^2)\right]\right\}\right] 
\nonumber\\ && \hspace{-22mm} \times \,
\exp \left[4\int_{\mu_F}^M \frac{d\mu}{\mu}\;
\gamma_{i/i}\left(\alpha_s(\mu)\right)\right] \;
\exp \left[2 d_{\alpha_s} \int_{\mu_R}^M \frac{d\mu}{\mu}\;
\beta\left(\alpha_s(\mu)\right)\right] \;
\nonumber\\ && \hspace{-22mm} \times \,
H\left(\alpha_s(\mu_R)\right) \;
S \left(\alpha_s(M/{\tilde N}) \right) \;
\exp \left[2\int_M^{M/{\tilde N}}
\frac{d\mu}{\mu}\; \Gamma_S
\left(\alpha_s(\mu)\right)\right]\!.
\label{resHS}
\eeqa

Factorization-derived resummation studies are based on the above formula 
or its 
equivalents. In particular, resummation studies for simple-color-structure 
processes such as Drell-Yan and Higgs production are based on \cite{GS,CT} 
while processes with more complicated color structure require the 
additional ingredients in \cite{KS,NKuni,NKNNNLO,KOS}.

The expressions for the quantities in the exponents  
up to three loops are given in Appendix A together with the necessary 
integrals for NNNLO calculations.
The first exponent includes corrections that are universal in hard 
scattering cross sections and only depends on the identity of the 
incoming partons (quarks or gluons) \cite{GS,CT}.
The second and third exponents control the factorization scale,
$\mu_F$, and the renormalization scale, $\mu_R$, dependence 
of the cross section, respectively.
$H$ is the hard-scattering function
for the scattering of partons, while $S$ is the
soft function describing noncollinear soft gluon emission \cite{KS}.
Also ${\tilde N}=N e^{\gamma_E}$ where $\gamma_E$ is the Euler constant.
The evolution of the soft function
follows from its renormalization group properties and
is given in terms of the soft anomalous dimension $\Gamma_S$
which can be explicitly derived through
the calculation of eikonal vertex corrections.

To include collinear singularities we essentially replace  
$-2(z^{N-1}-1)/(1-z)$ in the first exponent of Eq. (\ref{resHS}) by 
$2 z^{N-1}$. These corrections are of the form $\ln^m N/N$ in 
moment space (see Appendix A).
The leading collinear corrections come from the one-loop term in 
$A_i$ \cite{KLS,NKcoll}. However at NNLO and beyond one 
does not derive all the subleading terms from this exponential. We will 
show, nevertheless, that both analytically and numerically we get the 
dominant terms and the approximation is excellent.

The exponentials in the resummed cross section can be expanded
to any fixed order in the strong coupling $\alpha_s$ and then inverted 
to momentum space to provide explicit results for the higher-order corrections.
A fixed-order expansion avoids the problems with infrared singularities
in the exponents and thus no prescription is needed to deal with these
in our approach (see discussion in Ref. \cite{NKtop}).
The $n$-th order corrections can then be written as
\beq
{\hat \sigma}^{(n)}(z)=V^{(n)} \, \delta(1-z)+\sum_{k=0}^{2n-1} S_k^{(n)} 
\left[\frac{\ln^k(1-z)}{1-z}\right]_+ +\sum_{k=0}^{2n-1} C_k^{(n)} 
\ln^k(1-z) .
\label{sigmanp}
\eeq
Here $V^{(n)}$ are the virtual contributions, 
the soft contributions are of the form of plus distributions 
with coefficients $S_k^{(n)}$, and $C_k^{(n)}$ are the coefficients 
of the collinear logarithmic contributions. 

To calculate the hadronic cross section we convolute the partonic cross 
section with parton distribution functions $\phi$.
The $n$-th order corrections
to the hadronic cross section can then be written as 
\beq
\sigma^{(n)}=\tau \int_{\tau}^1 dx_1 \int_{\tau/x_1}^1 dx_2 
\int_0^1 dz \; 
\delta(\tau-x_1 x_2 z) \, \phi(x_1) \, \phi(x_2) \, {\hat \sigma}^{(n)}(z)\, .
\label{sigmanh}
\eeq
Here $\tau=Q^2/S$ and $z=Q^2/s$ for Drell-Yan production, with $S$ the 
center-of-mass energy squared of the incoming hadrons, while  
$\tau=m_H^2/S$ and $z=m_H^2/s$ for Higgs production.
Substituting the expression of Eq. (\ref{sigmanp}) in Eq. (\ref{sigmanh}) 
we have
\beqa
\sigma^{(n)}&=& \tau \int_{\tau}^1 dx_1 \int_{\tau/x_1}^1 dz \, 
\frac{1}{x_1 z} \, \phi(x_1) \, \phi\left(\frac{\tau}{x_1 z}\right) 
\left\{V^{(n)} \, \delta(1-z)+\sum_{k=0}^{2n-1} S_k^{(n)} 
\left[\frac{\ln^k(1-z)}{1-z}\right]_+ \right.
\nonumber \\ && \hspace{71mm} \left.
{}+\sum_{k=0}^{2n-1} C_k^{(n)} \ln^k(1-z) \right\}
\eeqa
which, after a few manipulations, gives
\beqa
\sigma^{(n)}&=& \tau \int_{\tau}^1 dx_1 \, \frac{1}{x_1} \, \phi(x_1)
\left\{\phi\left(\frac{\tau}{x_1}\right) 
\left[V^{(n)}+\sum_{k=0}^{2n-1} \frac{S_k^{(n)}}{k+1} 
\ln^{k+1}\left(1-\frac{\tau}{x_1}\right) \right]\right.
\nonumber \\ && \hspace{-20mm} \left. 
{}+\int_{\tau/x_1}^1 dz  \left[\frac{1}{z} 
\phi\left(\frac{\tau}{x_1 z}\right) -\phi\left(\frac{\tau}{x_1}\right)\right]
\sum_{k=0}^{2n-1} S_k^{(n)} \frac{\ln^k(1-z)}{1-z} 
+\int_{\tau/x_1}^1 dz \frac{1}{z} \phi\left(\frac{\tau}{x_1 z}\right)
\sum_{k=0}^{2n-1} C_k^{(n)} \ln^k(1-z) \right\} \, .
\nonumber \\
\eeqa

\mysection{Soft and collinear corrections for 
$b {\bar b} \rightarrow H$ through NNNLO}

In this section we calculate the complete soft-gluon corrections 
and a large class of collinear corrections beyond leading accuracy to the 
cross section for the process $b {\bar b} \rightarrow H$ through NNNLO.
In the calculation we also need the virtual corrections through NNLO 
\cite{HKbb,VR}.

We use the notation 
\beq
D_l(z)=\left[\frac{\ln^l(1-z)}{1-z}\right]_+
\eeq
to denote the plus distributions in the soft corrections.

The complete NLO soft and virtual corrections are then given by
\beqa
{\hat{\sigma}}^{(1)\, SV}_{b{\bar b} \rightarrow H} &=& 
F^B_{b{\bar b} \rightarrow H}(\mu_R^2) \frac{\alpha_s(\mu_R^2)}{\pi}
\left\{4C_F \, {\cal D}_1(z)  
-2 C_F \ln\left(\frac{\mu_F^2}{m_H^2}\right)\, {\cal D}_0(z) \right.
\nonumber \\ && \left.
{}+\left[C_F (-1+2\zeta_2) 
+\frac{3}{2} C_F \ln\left(\frac{\mu_R^2}{\mu_F^2}\right)
\right]\, \delta(1-z)\right\} 
\label{NLObbH}
\eeqa
where the color factors $C_F$, $C_A$ and the 
$\zeta_i$ constants are defined in Appendix A, and 
the Born term, $F^B_{b{\bar b} \rightarrow H}$ is given by  
\beq
F^B_{b{\bar b} \rightarrow H}(\mu_R^2)=\frac{\pi \lambda_b^2(\mu_R^2)}{12 m_H^2} \, .
\eeq
For the Standard Model process $\lambda_b(\mu_R^2)=\sqrt{2}m_b(\mu_R^2)/v$, 
where $v=246$ GeV is the Higgs boson vacuum expectation value 
and $m_b(\mu_R^2)$ is the ${\rm \overline{MS}}$ bottom 
quark running mass at scale $\mu_R$. 
In the MSSM the value of $\lambda_b$ 
depends on which supersymmetric Higgs boson is produced. In the numerical 
results below we will mostly show ratios of cross sections so the results 
are equally valid for the Standard Model and the MSSM. However, when 
values for cross sections are presented in this paper they are always for the 
Standard Model process. 

The leading and next-to-leading collinear corrections at NLO are 
\beq
{\hat{\sigma}}^{(1)\, C}_{b{\bar b} \rightarrow H}= 
F^B_{b{\bar b} \rightarrow H}(\mu_R^2) \frac{\alpha_s(\mu_R^2)}{\pi}
\left\{-4 C_F\, \ln(1-z) +2 C_F \ln\left(\frac{\mu_F^2}{m_H^2}\right)
+2 C_F\right\} \, . 
\eeq

The complete NNLO soft and virtual corrections are given by 
\beqa
{\hat{\sigma}}^{(2)\, SV}_{b{\bar b} \rightarrow H}&=&F^B_{b{\bar b} 
\rightarrow H}(\mu_R^2) \frac{\alpha_s^2(\mu_R^2)}{\pi^2} 
\left\{8 C_F^2 \, {\cal D}_3(z)
+\left[-\frac{11}{3}C_F C_A+\frac{2}{3}C_F n_f-12 C_F^2
\ln\left(\frac{\mu_F^2}{m_H^2}\right)\right] \, 
{\cal D}_2(z) \right.
\nonumber \\ && \hspace{-10mm}
{}+\left[-4 C_F^2(1+2\zeta_2)+\left(\frac{67}{9}-2\zeta_2\right)C_F C_A
-\frac{10}{9} C_F n_f+4 C_F^2\ln^2\left(\frac{\mu_F^2}{m_H^2}\right) \right.
\nonumber \\ && \hspace{-10mm} \quad \left.
{}-6 C_F^2 \ln\left(\frac{\mu_F^2}{m_H^2}\right) 
+C_F \left(6C_F+\frac{11}{3}C_A -\frac{2}{3}n_f\right)
\ln\left(\frac{\mu_R^2}{m_H^2}\right)\right] \,
{\cal D}_1(z)
\nonumber \\ && \hspace{-10mm}
{}+\left[16 \zeta_3 C_F^2+C_F C_A \left(-\frac{101}{27}+\frac{11}{3}\zeta_2
+\frac{7}{2}\zeta_3\right)+\frac{2}{3}C_F n_f\left(\frac{7}{9}-\zeta_2\right)
\right.
\nonumber \\ && \hspace{-10mm} \quad 
{}+C_F\left(3C_F+\frac{11}{12}C_A-\frac{n_f}{6}\right) 
\ln^2\left(\frac{\mu_F^2}{m_H^2}\right)
\nonumber \\ && \hspace{-10mm} \quad 
{}+\left((2+4\zeta_2)C_F^2+(-\frac{67}{18}+\zeta_2)C_F C_A
+\frac{5}{9}C_F n_f\right)  
\ln\left(\frac{\mu_F^2}{m_H^2}\right) 
\nonumber \\ && \hspace{-10mm} \quad \left. 
{}-C_F\left(3C_F+\frac{11}{6}C_A-\frac{n_f}{3}\right)
\ln\left(\frac{\mu_F^2}{m_H^2}\right)
\ln\left(\frac{\mu_R^2}{m_H^2}\right) \right] \, 
{\cal D}_0(z) 
\nonumber \\ && \hspace{-10mm}
{}+\left[C_F^2\left(1-\frac{15}{4}\zeta_3+\frac{\zeta_2^2}{10}\right)
+C_F C_A \left(\frac{83}{72}+\frac{29}{18}\zeta_2-\frac{\zeta_3}{2}
-\frac{3}{20} \zeta_2^2\right) +n_f C_F\left(\frac{1}{18}
-\frac{5}{18}\zeta_2+\frac{\zeta_3}{2}\right) \right.
\nonumber \\ && \hspace{-10mm} \quad 
{}+\left(C_F^2(\frac{9}{8}-2\zeta_2)+\frac{11}{16}C_F C_A
-\frac{1}{8} C_F n_f\right) \ln^2\left(\frac{\mu_F^2}{m_H^2}\right) 
\nonumber \\ && \hspace{-10mm} \quad 
{}+\left(C_F^2(\frac{21}{16}-\frac{3}{2}\zeta_2-11\zeta_3)
+C_F C_A (-\frac{17}{48}-\frac{11}{6}\zeta_2+\frac{3}{2}\zeta_3)
+\frac{1}{3} C_F n_f (\frac{1}{8}+\zeta_2)\right)
\ln\left(\frac{\mu_F^2}{m_H^2}\right) 
\nonumber \\ && \hspace{-10mm} \quad 
{}+\left(-\frac{9}{4}C_F^2-\frac{11}{8}C_F C_A+\frac{1}{4} C_F n_f\right)
\ln\left(\frac{\mu_F^2}{m_H^2}\right) \ln\left(\frac{\mu_R^2}{m_H^2}\right)
\nonumber \\ && \hspace{-10mm} \quad 
{}+\left(\frac{9}{8} C_F^2+\frac{11}{16} C_F C_A-\frac{n_f}{8} C_F\right)
\ln^2\left(\frac{\mu_R^2}{m_H^2}\right)
\nonumber \\ && \hspace{-10mm} \quad \left. \left.
{}+\left(C_F^2\left(-\frac{21}{16}+3\zeta_2\right)
+C_F C_A\left(\frac{53}{48}+\frac{11}{6}\zeta_2\right)
+C_F n_f \left(-\frac{1}{24}-\frac{\zeta_2}{3}\right)
\right)\ln\left(\frac{\mu_R^2}{m_H^2}\right)
\right] \delta(1-z)\right\} \, , 
\nonumber \\
\label{NNLObbH}
\eeqa
where
$n_f$ is the number of light quark flavors.
This is in agreement with Refs. \cite{HKbb,VR}. 

It is interesting to note here that the scale-independent contribution 
in the $D_0$ terms is solely due to $\zeta_3$ terms and the two-loop function 
$G_{q{\bar q}}^{(2)}$ defined in Appendix A. It was shown in Ref. 
\cite{NKRVtop} that for top quark production the inclusion of $\zeta_i$ 
and two-loop $G_{q{\bar q}}^{(2)}$ terms in the $D_0$ coefficient 
for that process provided the 
majority of the $D_0$ corrections, even though the complete two-loop $D_0$ 
terms are not known for that process. This was demonstrated by comparing 
the corrections in two different kinematics formulations for $t{\bar t}$
production and showing that the results agree only with the inclusion of 
these $\zeta_i$ and two-loop terms.

The leading and some subleading collinear corrections at NNLO are
\beqa
{\hat{\sigma}}^{(2)\, C}_{b{\bar b} \rightarrow H}&=&
F^B_{b{\bar b} \rightarrow H}(\mu_R^2) \frac{\alpha_s^2(\mu_R^2)}{\pi^2} 
\left\{-8 C_F^2 \, \ln^3(1-z) \right.
\nonumber \\ && \hspace{-10mm}
{}+\left[12C_F^2+\frac{11}{3}C_F C_A-\frac{2}{3}C_F n_f+12 C_F^2
\ln\left(\frac{\mu_F^2}{m_H^2}\right)\right] \, \ln^2(1-z) 
\nonumber \\ && \hspace{-10mm}
{}+\left[4C_F^2(1+2\zeta_2)+C_F C_A\left(2\zeta_2-\frac{100}{9}\right)
+\frac{16}{9}n_f C_F -4 C_F^2\ln^2\left(\frac{\mu_F^2}{m_H^2}\right)  \right.
\nonumber \\ && \hspace{-5mm} \left.
{} -C_F\left(\frac{11}{3}C_A -\frac{2}{3}n_f\right)
\ln\left(\frac{\mu_R^2}{m_H^2}\right)
-6 C_F^2 \ln\left(\frac{\mu_F^2}{m_H^2}\right) 
-6 C_F^2 \ln\left(\frac{\mu_R^2}{m_H^2}\right) \right] \, \ln(1-z)
\nonumber \\ && \hspace{-10mm} 
{}-C_F\left(C_F+\frac{11}{12}C_A-\frac{n_f}{6}\right)
\ln^2\left(\frac{\mu_F^2}{m_H^2}\right)
+C_F\left(3C_F+\frac{11}{6}C_A-\frac{1}{3}n_f\right)
\ln\left(\frac{\mu_F^2}{m_H^2}\right)
\ln\left(\frac{\mu_R^2}{m_H^2}\right)
\nonumber \\ && \hspace{-10mm} 
{}+\left(-5 C_F^2+\frac{67}{18}C_F C_A-4C_F^2 \zeta_2-C_F C_A \zeta_2
-\frac{5}{9} n_f C_F\right)\ln\left(\frac{\mu_F^2}{m_H^2}\right)
\nonumber \\ && \hspace{-10mm} \left.
{}+C_F\left(3C_F+\frac{11}{6}C_A -\frac{1}{3}n_f\right)
\ln\left(\frac{\mu_R^2}{m_H^2}\right) 
\right\} \, . 
\label{NNLOcbbh}
\eeqa
Several remarks are in order here. The leading collinear (LC) logarithms 
at NNLO (i.e. $\ln^3(1-z)$) are complete. The next-to-leading collinear 
(NLC) logarithms (i.e $\ln^2(1-z)$) are not complete. However, numerically 
they are an 
excellent approximation to the complete NLC terms. Also analytically 
the $C_F C_A$, $n_f C_F$ and the $\ln(\mu_F^2/m_H^2)$ terms are exact. 
The next-to-next-to-leading collinear (NNLC) logarithms 
(i.e $\ln(1-z)$) are also not complete. However, again  numerically they are 
an excellent approximation to the complete NNLC terms. Also analytically the 
$C_F^2 \zeta_2$, $C_F C_A \zeta_2$, $\ln^2(\mu_F^2/m_H^2)$, and 
$\ln(\mu_R^2/m_H^2)$ terms are exact. Finally, in the constant terms we 
only show the scale-dependent terms; the $\ln^2(\mu_F^2/m_H^2)$, 
$\ln(\mu_F^2/m_H^2) \ln(\mu_R^2/m_H^2)$ and $\ln(\mu_R^2/m_H^2)$ terms are 
exact, while the $\ln(\mu_F^2/m_H^2)$ terms are almost exact. More details on 
the numerical approximation are given in Section 4, and on the 
analytical approximation in the context of the related Drell-Yan process, 
where explicit results involving the color factors have been published 
\cite{DY}, in Appendix B.

The complete NNNLO soft-gluon corrections are given by
\beqa
{\hat{\sigma}}^{(3)\, S}_{b{\bar b} \rightarrow H}&=& 
F^B_{b{\bar b} \rightarrow H}(\mu_R^2) \frac{\alpha_s^3(\mu_R^2)}{\pi^3} \;  
\left \{8 C_F^3 \; {\cal D}_5(z)
+\left[-\frac{110}{9} C_F^2 C_A+\frac{20}{9} C_F^2 n_f 
-20 C_F^3 \ln\left(\frac{\mu_F^2}{m_H^2}\right) \right] \; {\cal D}_4(z) 
\right.
\nonumber \\ && \hspace{-15mm}
{}+\left[-8C_F^3 (1+6 \zeta_2) + C_F^2 C_A \left(\frac{268}{9}-8\zeta_2\right)
+\frac{121}{27} C_F C_A^2-\frac{40}{9}C_F^2 n_f
-\frac{44}{27} C_F C_A n_f+\frac{4}{27} C_F n_f^2\right.
\nonumber \\ && \hspace{-10mm}
{}+16 C_F^3 \ln^2\left(\frac{\mu_F^2}{m_H^2}\right)
+\left(-12 C_F^3+\frac{88}{9} C_F^2 C_A -\frac{16}{9} C_F^2 n_f\right) 
\ln\left(\frac{\mu_F^2}{m_H^2}\right) 
\nonumber \\ && \hspace{-10mm}\left.
{}+\left(12C_F^3+\frac{44}{3} C_F^2 C_A -\frac{8}{3} C_F^2 n_f\right) 
\ln\left(\frac{\mu_R^2}{m_H^2}\right)\right] \; {\cal D}_3(z) 
\nonumber \\ && \hspace{-15mm}
{}+\left[160 \zeta_3 C_F^3+C_F^2 C_A\left(-\frac{169}{9}+\frac{176}{3}\zeta_2
+21 \zeta_3\right)+C_F C_A^2\left(-\frac{445}{27}+\frac{11}{3}\zeta_2\right)
+C_F^2 n_f\left(\frac{53}{18}-\frac{32}{3}\zeta_2\right)
\right.
\nonumber \\ && \hspace{-10mm}
{}+C_F C_A n_f\left(\frac{289}{54}-\frac{2}{3}\zeta_2\right)
-\frac{10}{27}C_F n_f^2-4C_F^3\ln^3\left(\frac{\mu_F^2}{m_H^2}\right)
+C_F^2\left(18 C_F+\frac{11}{2}C_A-n_f\right) 
\ln^2\left(\frac{\mu_F^2}{m_H^2}\right)
\nonumber \\ && \hspace{-10mm}
{}+\left(12 (1+6\zeta_2)C_F^3+(-\frac{235}{6}+12\zeta_2)C_F^2 C_A
+\frac{17}{3}C_F^2 n_f\right) \ln\left(\frac{\mu_F^2}{m_H^2}\right)
\nonumber \\ && \hspace{-10mm} 
{}+(-18C_F^3-22 C_F^2 C_A+4C_F^2 n_f) \ln\left(\frac{\mu_F^2}{m_H^2}\right)
\ln\left(\frac{\mu_R^2}{m_H^2}\right)
\nonumber \\ && \hspace{-10mm} \left.
{}+\left(-\frac{11}{2}C_F^2C_A-\frac{121}{18} C_F C_A^2+C_F^2 n_f
+\frac{22}{9} C_F C_A n_f-\frac{2}{9} C_F n_f^2\right)
\ln\left(\frac{\mu_R^2}{m_H^2}\right)\right] {\cal D}_2(z) 
\nonumber \\ && \hspace{-15mm}
{}+\left[C_F^3\left(4+16\zeta_2-15\zeta_3
-192\zeta_4+\frac{162}{5}\zeta_2^2\right)
+C_F^2 C_A\left(-\frac{17}{6}-\frac{326}{9}\zeta_2-90\zeta_3
+\frac{57}{5}\zeta_2^2\right)\right.
\nonumber \\ && \hspace{-10mm}
{}+C_F C_A^2\left(\frac{15503}{648}-\frac{188}{9}\zeta_2-11\zeta_3
+\frac{11}{5}\zeta_2^2\right)
+C_F^2 n_f\left(-\frac{23}{24}+\frac{50}{9}\zeta_2+20\zeta_3\right)
\nonumber \\ && \hspace{-10mm}
{}+C_F C_A n_f\left(-\frac{2051}{324}+6\zeta_2\right)
+C_F n_f^2\left(\frac{25}{81}-\frac{4}{9}\zeta_2\right)
\nonumber \\ && \hspace{-10mm}
{}+\left(-6C_F^3-\frac{11}{3}C_F^2 C_A+\frac{2}{3} C_F^2 n_f\right)
\ln^3\left(\frac{\mu_F^2}{m_H^2}\right)
\nonumber \\ && \hspace{-10mm}
{}+\left((\frac{1}{2}-32\zeta_2)C_F^3+(\frac{635}{36}-4\zeta_2)C_F^2 C_A
-\frac{49}{18} C_F^2 n_f\right)\ln^2\left(\frac{\mu_F^2}{m_H^2}\right)
\nonumber \\ && \hspace{-10mm}
{}+C_F^2\left(6C_F+\frac{22}{3}C_A-\frac{4}{3}n_f\right)
\ln^2\left(\frac{\mu_F^2}{m_H^2}\right)
\ln\left(\frac{\mu_R^2}{m_H^2}\right)
\nonumber \\ && \hspace{-10mm}
{}+C_F^2(-9C_F-11C_A+2n_f)\ln\left(\frac{\mu_F^2}{m_H^2}\right)
\ln\left(\frac{\mu_R^2}{m_H^2}\right)
\nonumber \\ && \hspace{-10mm}
{}+\left((\frac{21}{4}+18\zeta_2-172 \zeta_3) C_F^3
+(\frac{257}{108}-\frac{101}{3}\zeta_2-8\zeta_3)C_F^2 C_A
+(-\frac{13}{54}+\frac{20}{3}\zeta_2)C_F^2 n_f\right)
\ln\left(\frac{\mu_F^2}{m_H^2}\right)
\nonumber \\ && \hspace{-10mm}
{}+\left(\frac{9}{2}C_F^3+\frac{33}{4}C_F^2C_A+\frac{121}{36}C_F C_A^2
-\frac{3}{2}n_f C_F^2-\frac{11}{9} C_F C_A n_f
+\frac{1}{9} C_F n_f^2\right)\ln^2\left(\frac{\mu_R^2}{m_H^2}\right)
\nonumber \\ && \hspace{-10mm}
{}+\left(C_F^3\left(-\frac{21}{4}-12\zeta_2\right)
+\left(\frac{143}{12}-\frac{53}{3}\zeta_2\right) C_F^2 C_A
+(\frac{445}{27}-\frac{11}{3}\zeta_2)C_F C_A^2
+\frac{1}{3}(-5+8\zeta_2) C_F^2 n_f \right.
\nonumber \\ && \left. \left. 
{}+(-\frac{289}{54}+\frac{2}{3}\zeta_2)C_F C_A n_f 
+\frac{10}{27} C_F n_f^2\right)\ln\left(\frac{\mu_R^2}{m_H^2}\right)\right] 
\; {\cal D}_1(z) 
\nonumber \\ && \hspace{-15mm}
{}+\left[C_F^3(-16\zeta_3+192\zeta_5-96 \zeta_2 \zeta_3)
+C_F^2 C_A\left(\frac{101}{27}+\frac{103}{27}\zeta_2+\frac{1009}{18}\zeta_3
+\frac{220}{3}\zeta_4-22\zeta_2^2-23\zeta_2\zeta_3\right) \right.
\nonumber \\ && \hspace{-10mm}
{}+C_F C_A^2 \left(-\frac{297029}{23328}+\frac{6139}{324}\zeta_2
+\frac{2509}{108}\zeta_3
-6\zeta_5-\frac{187}{60}\zeta_2^2-\frac{11}{6}\zeta_2\zeta_3\right) 
\nonumber \\ && \hspace{-10mm}
{}+C_F^2 n_f \left(\frac{421}{288}-\frac{47}{54}\zeta_2-\frac{179}{18}\zeta_3
-\frac{40}{3} \zeta_4 +\frac{19}{5}\zeta_2^2\right) 
+C_F C_A n_f \left(\frac{31313}{11664}-\frac{1837}{324}\zeta_2
-\frac{155}{36}\zeta_3+\frac{23}{30}\zeta_2^2\right) 
\nonumber \\ && \hspace{-10mm}
{}+C_F n_f^2 \left(-\frac{58}{729}+\frac{10}{27}\zeta_2
+\frac{5}{27}\zeta_3\right) 
\nonumber \\ && \hspace{-10mm}
{}+\left((-\frac{9}{4}+4\zeta_2) C_F^3-\frac{11}{4}C_F^2 C_A
-\frac{121}{216}C_F C_A^2+\frac{1}{2}C_F^2 n_f+\frac{11}{54} C_F C_A n_f
-\frac{1}{54}C_F n_f^2\right)\ln^3\left(\frac{\mu_F^2}{m_H^2}\right)
\nonumber \\ && \hspace{-10mm}
{}+\left((-\frac{21}{8}-9\zeta_2+38 \zeta_3) C_F^3
+(\frac{43}{8}+\frac{\zeta_2}{3}-3\zeta_3)C_F^2 C_A
+(\frac{445}{108}-\frac{11}{12}\zeta_2)C_F C_A^2 \right. 
\nonumber \\ && \hspace{-10mm} \left.
{}+(-\frac{7}{8}-\frac{\zeta_2}{3}) C_F^2 n_f
+(-\frac{289}{216}+\frac{\zeta_2}{6}) C_F C_A n_f
+\frac{5}{54} C_F n_f^2\right)
\ln^2\left(\frac{\mu_F^2}{m_H^2}\right)
\nonumber \\ && \hspace{-10mm}
{}+\left(\frac{9}{2}C_F^3+\frac{55}{8} C_F^2 C_A +\frac{121}{72}C_F C_A^2 
-\frac{5}{4}C_F^2 n_f-\frac{11}{18} C_F C_A n_f+\frac{1}{18} C_F n_f^2\right)
\ln^2\left(\frac{\mu_F^2}{m_H^2}\right)\ln\left(\frac{\mu_R^2}{m_H^2}\right)
\nonumber \\ && \hspace{-10mm}
{}+\left(-\frac{9}{4}C_F^3-\frac{33}{8}C_F^2C_A-\frac{121}{72} C_F C_A^2
+\frac{3}{4}n_f C_F^2+\frac{11}{18} C_F C_A n_f
-\frac{1}{18} C_F n_f^2 \right)
\ln\left(\frac{\mu_F^2}{m_H^2}\right)\ln^2\left(\frac{\mu_R^2}{m_H^2}\right)
\nonumber \\ && \hspace{-10mm}
{}+\left(C_F^3\left(\frac{21}{8}+6\zeta_2\right)+(-\frac{143}{24}
+\frac{53}{6}\zeta_2)C_F^2 C_A
+(-\frac{445}{54}+\frac{11}{6}\zeta_2) C_F C_A^2
+(\frac{5}{6}-\frac{4}{3}\zeta_2)C_F^2 n_f \right.
\nonumber \\ && \left. 
+(\frac{289}{108}-\frac{\zeta_2}{3}) C_F C_A n_f
-\frac{5}{27} C_F n_f^2 \right)
\ln\left(\frac{\mu_F^2}{m_H^2}\right)\ln\left(\frac{\mu_R^2}{m_H^2}\right)
\nonumber \\ && \hspace{-10mm}
{}+\left((-2-8\zeta_2-\frac{33}{2} \zeta_3
+96\zeta_4-\frac{81}{5}\zeta_2^2) C_F^3 
+(\frac{253}{36}+\frac{227}{18}\zeta_2+\frac{125}{12}\zeta_3
-\frac{57}{10}\zeta_2^2) C_F^2 C_A \right.
\nonumber \\ &&
{}+(-\frac{245}{48}+\frac{67}{18}\zeta_2-\frac{11}{12}\zeta_3
-\frac{11}{10}\zeta_2^2)C_F C_A^2
+(-\frac{43}{144}-\frac{16}{9}\zeta_2-\frac{14}{3}\zeta_3) C_F^2 n_f
\nonumber \\ && \left.
{}+(\frac{209}{216}-\frac{5}{9}\zeta_2+\frac{7}{6}\zeta_3)C_F C_A n_f
+\frac{1}{54} C_F n_f^2 \right) 
\ln\left(\frac{\mu_F^2}{m_H^2}\right)
\nonumber \\ && \hspace{-10mm}
{}+\left(24 \zeta_3 C_F^3+\left(-\frac{101}{18}+\frac{11}{2}\zeta_2
+\frac{415}{12} \zeta_3\right) C_F^2 C_A 
+(-\frac{1111}{162}+\frac{121}{18}\zeta_2+\frac{77}{12}\zeta_3) C_F C_A^2
\right.
\nonumber \\ && 
{}+\left(\frac{7}{9}-\zeta_2-\frac{16}{3}\zeta_3\right) C_F^2 n_f
+(\frac{178}{81}-\frac{22}{9}\zeta_2-\frac{7}{6} \zeta_3) C_F C_A n_f 
\nonumber \\ &&  \left. \left. \left.
{}+(-\frac{14}{81}+\frac{2}{9}\zeta_2) C_F n_f^2
\right) \ln\left(\frac{\mu_R^2}{m_H^2}\right)\right]\; {\cal D}_0(z) \right\}
\, .
\label{NNNLObbH}
\eeqa
This is in agreement with Ref. \cite{VR}.

The leading and some subleading collinear corrections at NNNLO are
\beqa
{\hat{\sigma}}^{(3)\, C}_{b{\bar b} \rightarrow H}&=& 
F^B_{b{\bar b} \rightarrow H}(\mu_R^2) \frac{\alpha_s^3(\mu_R^2)}{\pi^3} \,  
\left \{-8 C_F^3 \, \ln^5(1-z) \right.
\nonumber \\ && \hspace{-15mm}  
{}+\left[\frac{110}{9} C_F^2 C_A-\frac{20}{9} C_F^2 n_f 
+20 C_F^3 \ln\left(\frac{\mu_F^2}{m_H^2}\right) \right] \, \ln^4(1-z) 
\nonumber \\ && \hspace{-15mm}  
{}+\left[8 C_F^3(1+6\zeta_2)+C_F^2 C_A\left(-\frac{488}{9}+8\zeta_2\right)
-\frac{121}{27} C_F C_A^2+C_F\frac{n_f}{27}(240C_F+44 C_A-4n_f) \right. 
\nonumber \\ && \hspace{-10mm}   
{}-16 C_F^3 \ln^2\left(\frac{\mu_F^2}{m_H^2}\right)
+C_F^2\left(12 C_F-\frac{88}{9} C_A+\frac{16}{9}n_f\right)
\ln\left(\frac{\mu_F^2}{m_H^2}\right)
\nonumber \\ && \hspace{-10mm} \left.  
{}+\left(-12C_F^3-\frac{44}{3} C_F^2 C_A +\frac{8}{3} C_F^2 n_f\right) 
\ln\left(\frac{\mu_R^2}{m_H^2}\right)\right] \, \ln^3(1-z) 
\nonumber \\ && \hspace{-15mm} \left.
{}+\left[4 C_F^3 \ln^3\left(\frac{\mu_F^2}{m_H^2}\right)
+C_F^2 (18C_F+22 C_A -4 n_f) \ln\left(\frac{\mu_F^2}{m_H^2}\right)
\ln\left(\frac{\mu_R^2}{m_H^2}\right)\right] \, \ln^2(1-z) \right\} \, . 
\eeqa
Again we note that only the LC ($\ln^5(1-z)$) terms are complete. The 
NLC ($\ln^4(1-z)$) and NNLC ($\ln^3(1-z)$) terms 
are not complete but, based on our study at NNLO, we expect them to be 
a very good approximation to the complete terms both in their analytical 
structure and numerically. Finally, in the $\ln^2(1-z)$ terms we only 
show the scale terms that we expect to be exact at this accuracy. 

\mysection{Cross sections for $b {\bar b} \rightarrow H$ at the Tevatron 
and the LHC}

We now present a numerical study of the contribution of the corrections 
to the cross section for $b {\bar b} \rightarrow H$ at the Tevatron 
and the LHC. We use the bottom quark parton distribution functions (pdf) from 
the MRST2006 NNLO set of parton densities \cite{MRST2006}.
We are interested in the total cross section and the relative size of the 
higher-order contributions to it (the effect of parton radiation in transverse
momentum distributions for this process has been studied in \cite{BNY}).
We also provide a few results for $gg \rightarrow H$ using the gluon pdf from 
\cite{MRST2006} to compare with $b {\bar b} \rightarrow H$.
In the results below we set the factorization and renormalization scales 
equal to each other and denote this common scale by $\mu$.

\begin{figure}
\begin{center}
\hspace{-5mm}
\includegraphics[width=8cm]{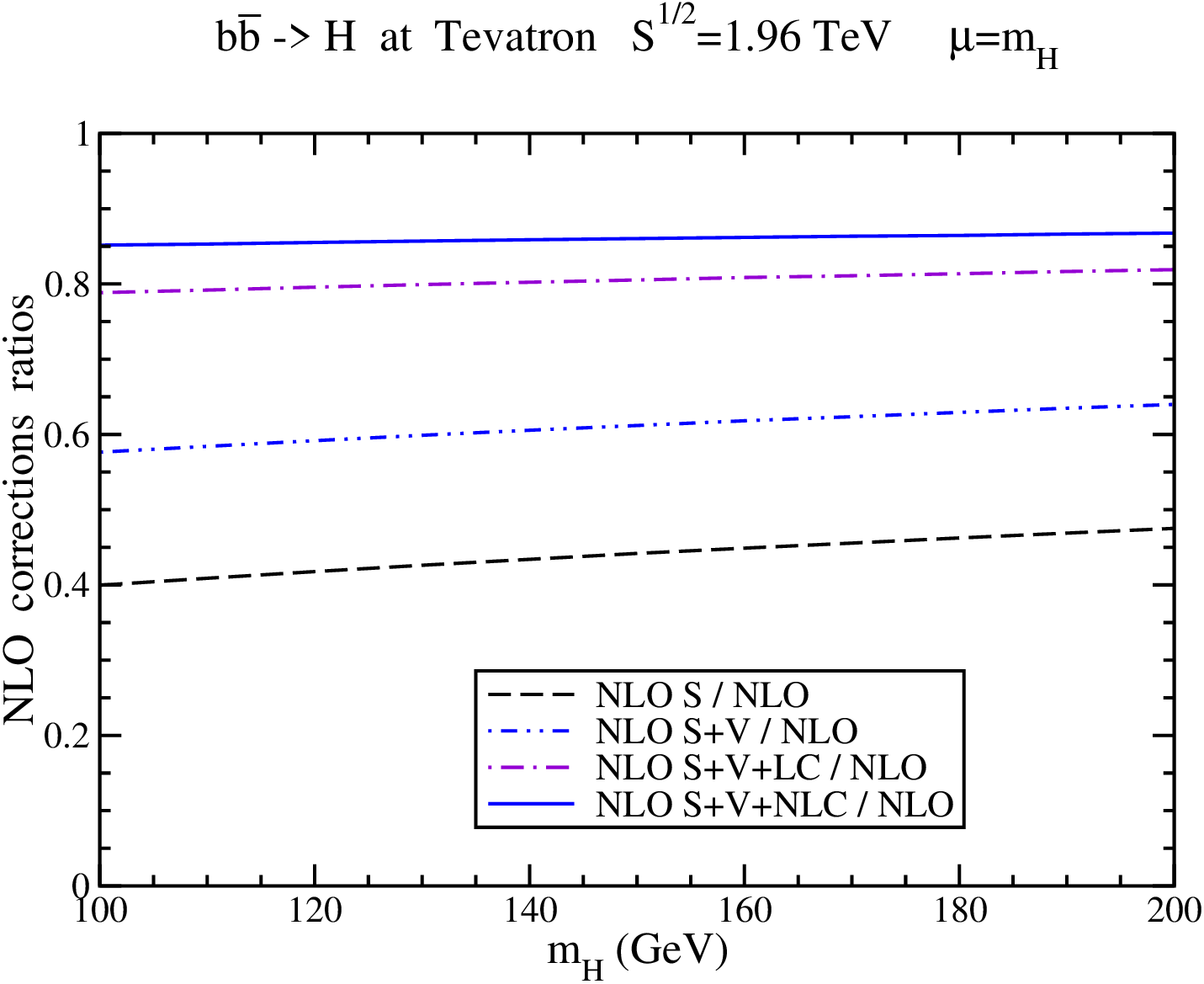}
\hspace{5mm}
\includegraphics[width=8cm]{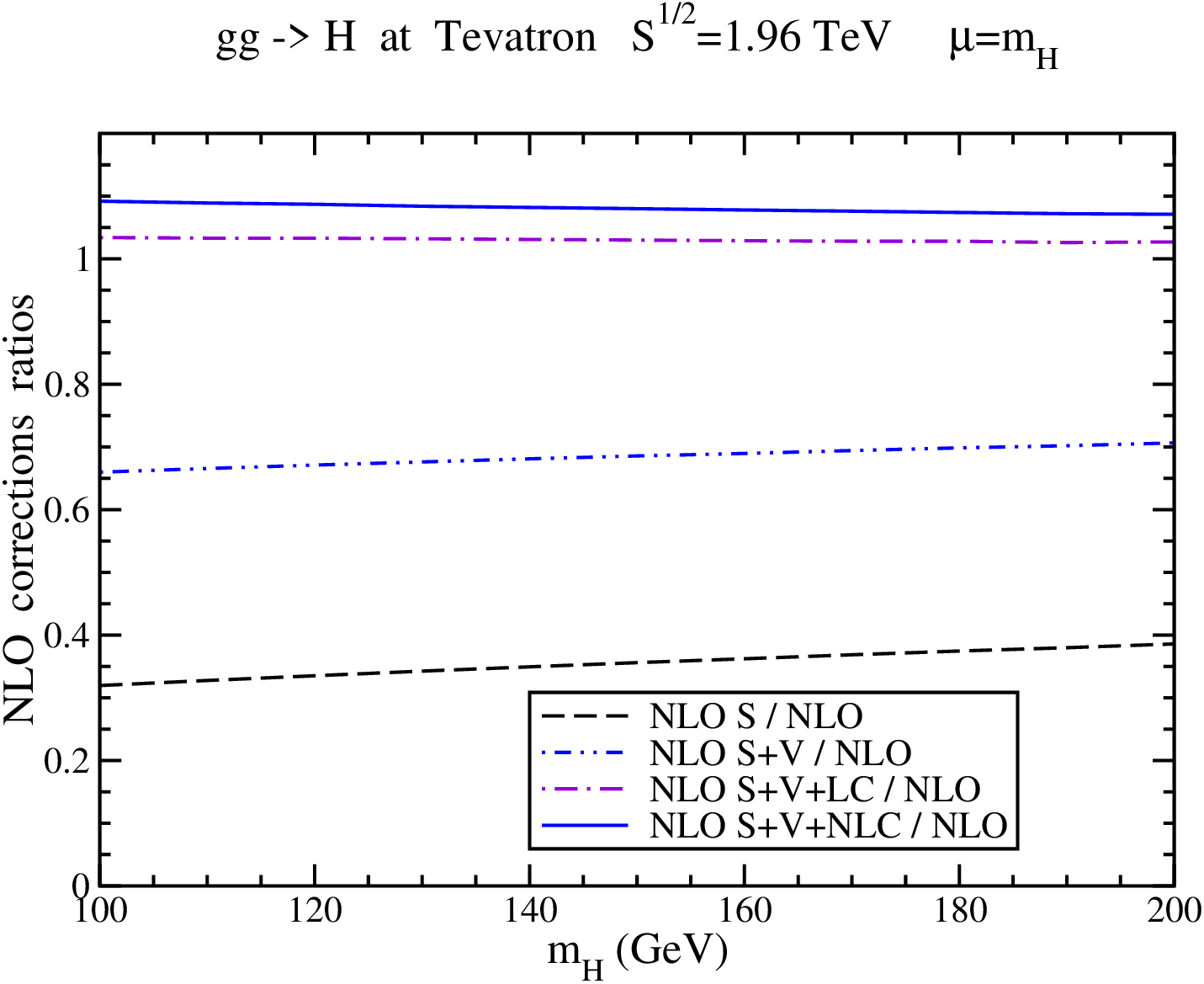}
\caption{Left: The NLO ratios for $b {\bar b} \rightarrow H$ at the Tevatron.
Here $\mu=\mu_F=\mu_R=m_H$. Right: The NLO ratios for $gg \rightarrow H$ 
at the Tevatron.}
\label{NLOtevplot}
\end{center}
\end{figure}

In Figure 1 we investigate the contribution of various terms 
to the complete NLO corrections for Higgs production at the 
Tevatron, with $\sqrt{S}=1.96$ TeV and setting $\mu=m_H$, 
by plotting the corresponding ratios. 
It is important to clarify that in this figure NLO denotes 
the ${\cal O}(\alpha_s)$ 
corrections only (i.e. without the Born term). The left-hand side shows 
results for the process $b {\bar b} \rightarrow H$ and the right-hand 
side shows for comparison results for $gg \rightarrow H$. The curve 
marked NLO S / NLO denotes the percentage contribution of the NLO soft 
(S) corrections to the total NLO corrections. We see that this contribution 
does not surpass 50\% for $b {\bar b} \rightarrow H$ and 40\% for 
$gg \rightarrow H$ and thus the soft-gluon approximation is by itself 
inadequate. Adding the virtual terms to the soft, we get the soft plus 
virtual (S+V) approximation which, although better than the soft approximation 
alone, still does not provide a good approximation of the full corrections. 
Adding collinear corrections 
clearly substantially improves the situation. Simply by adding the 
leading collinear (LC) logarithms to the soft and virtual terms, the resulting 
S+V+LC approximation accounts for about 80\% of the total NLO corrections 
for  $b {\bar b} \rightarrow H$. For $gg \rightarrow H$ the S+V+LC 
approximation overestimates the total NLO corrections by a few percent. 
If we further add the next-to-leading collinear terms the approximation 
(S+V+NLC) gets even better for $b {\bar b} \rightarrow H$, reaching over 
85\% of the total corrections. For $gg \rightarrow H$ the inclusion of NLC 
terms overestimates the cross section, but still by less than 10\%. 
Clearly the inclusion of collinear terms greatly improves the approximation 
in both cases, and in particular for $b {\bar b} \rightarrow H$ it is 
important to include the NLC terms.

\begin{figure}
\begin{center}
\hspace{-5mm}
\includegraphics[width=8cm]{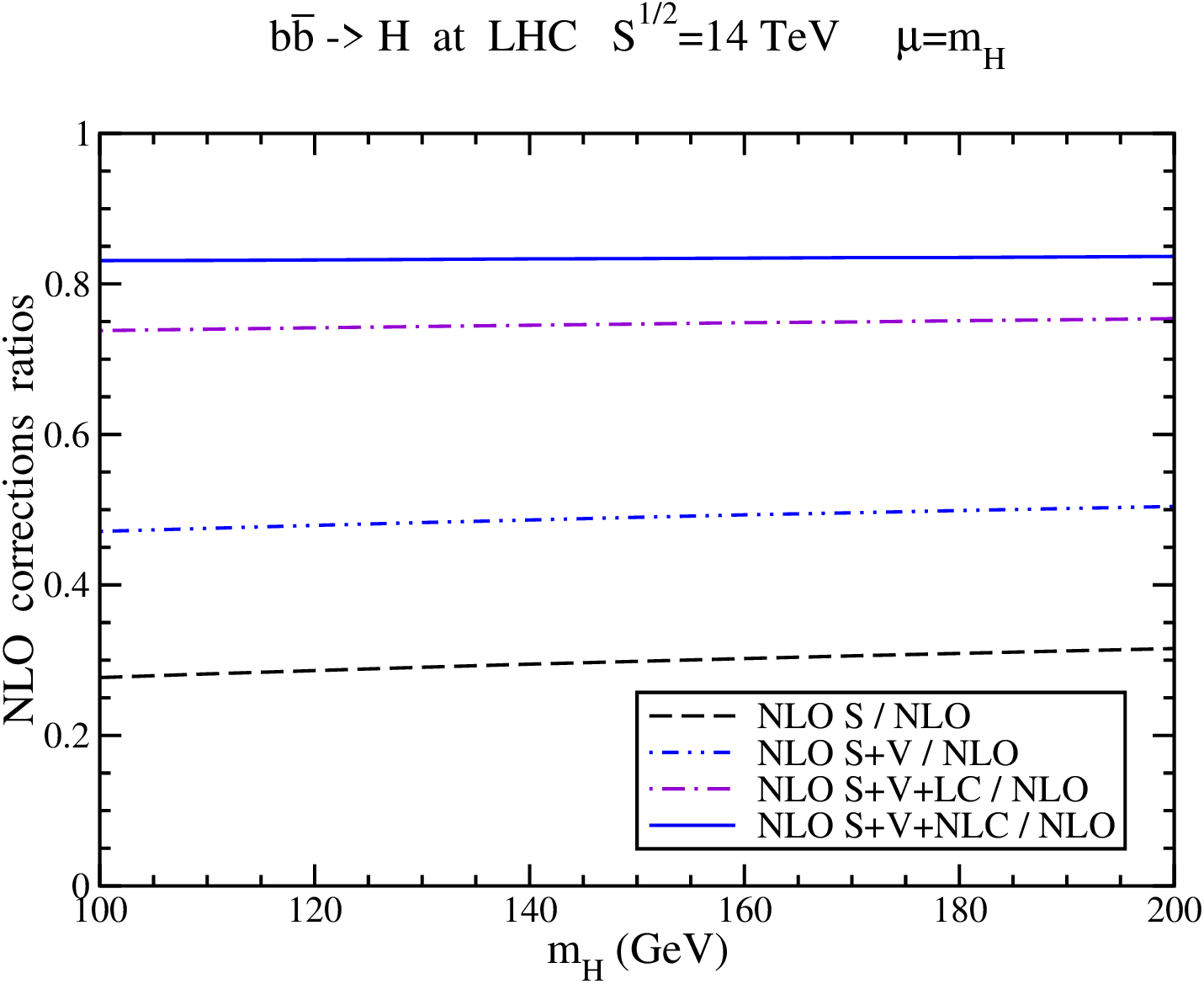}
\hspace{5mm}
\includegraphics[width=8cm]{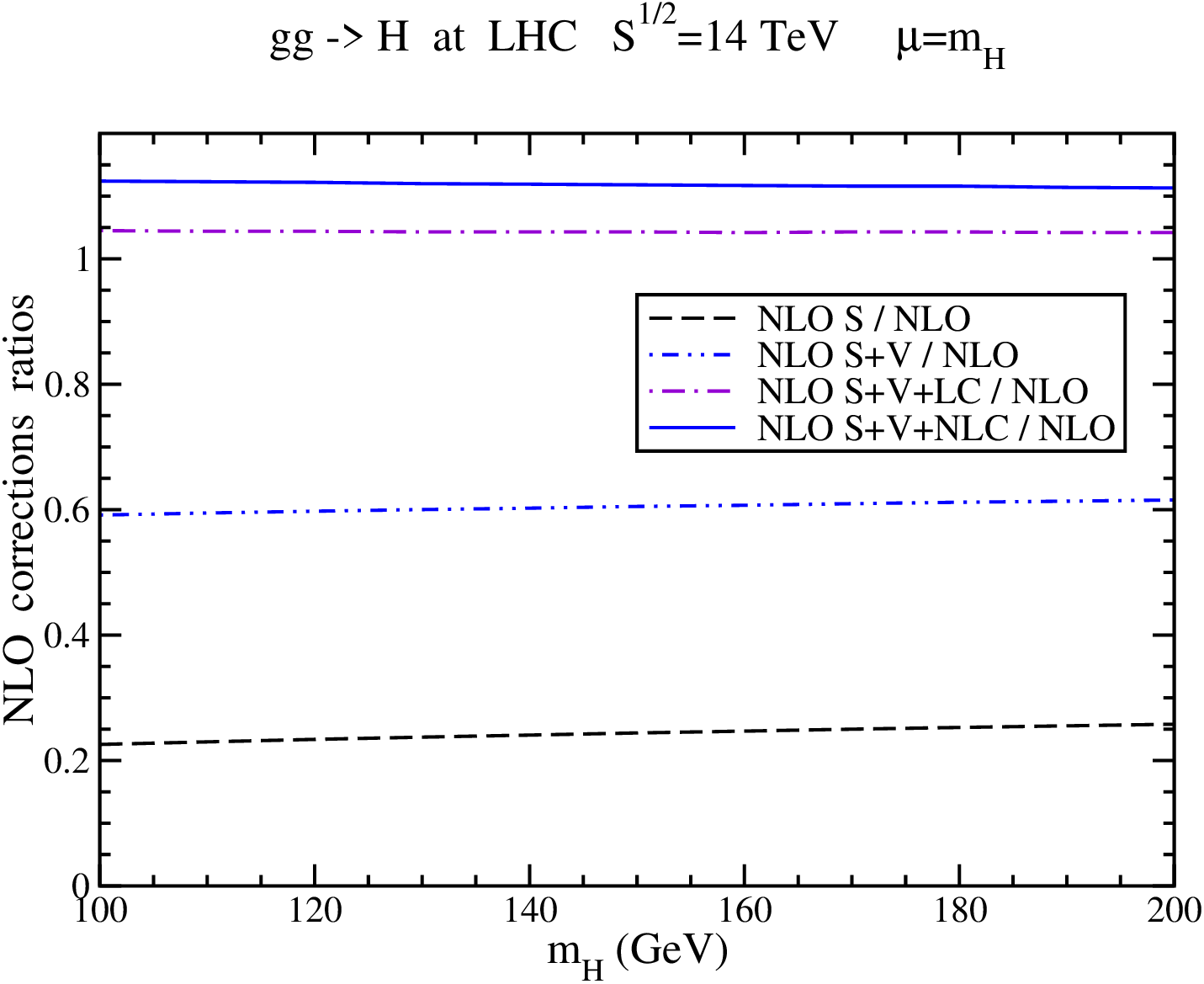}
\caption{Left: The NLO ratios for $b {\bar b} \rightarrow H$ at the LHC.
Here $\mu=\mu_F=\mu_R=m_H$. Right: The NLO ratios for $gg \rightarrow H$ 
at the LHC.}
\label{NLOlhcplot}
\end{center}
\end{figure}

Figure 2 shows the corresponding NLO ratios for Higgs production at the 
LHC, $\sqrt{S}=14$ TeV. Again, the left-hand side shows 
results for the process $b {\bar b} \rightarrow H$ and the right-hand 
side shows for comparison results for $gg \rightarrow H$.
We note that the results are quite similar to those for the Tevatron for 
both processes and our observations and conclusions are the same. 
In fact the soft and the S+V approximations are even more inadequate at 
the LHC (consistent with the fact that we are further away from threshold) 
which makes it even more essential to include the collinear corrections.
We also note that in both Figure 1 and Figure 2 our results for 
$gg \rightarrow H$ are consistent 
with those published in \cite{RSvN}. Since $gg \rightarrow H$ has been 
studied at length elsewhere with results consistent with ours, 
we focus the rest of the presentation of our numerical results 
solely on the process $b {\bar b} \rightarrow H$. 

\begin{figure}
\begin{center}
\hspace{-5mm}
\includegraphics[width=8cm]{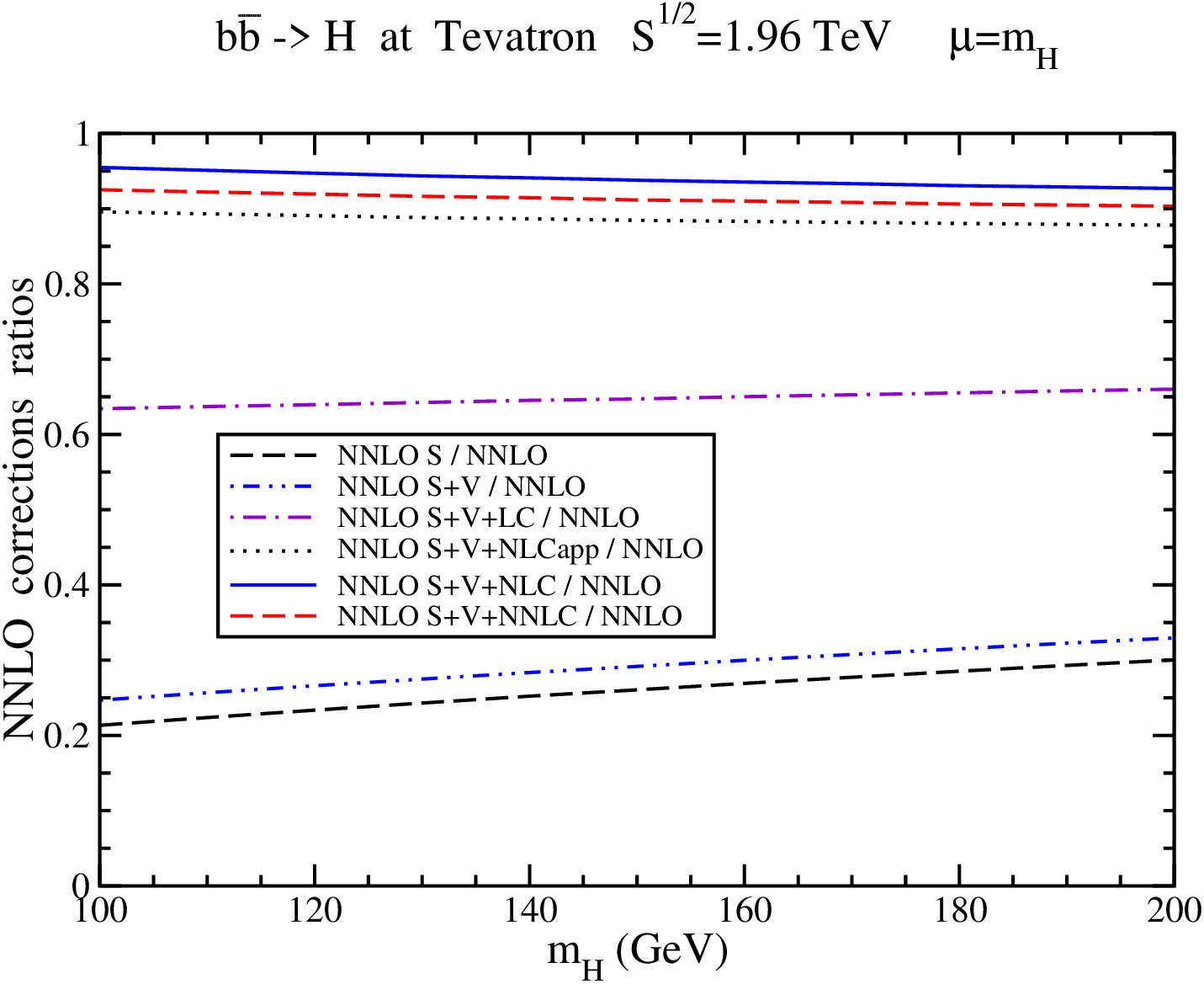}
\hspace{5mm}
\includegraphics[width=8cm]{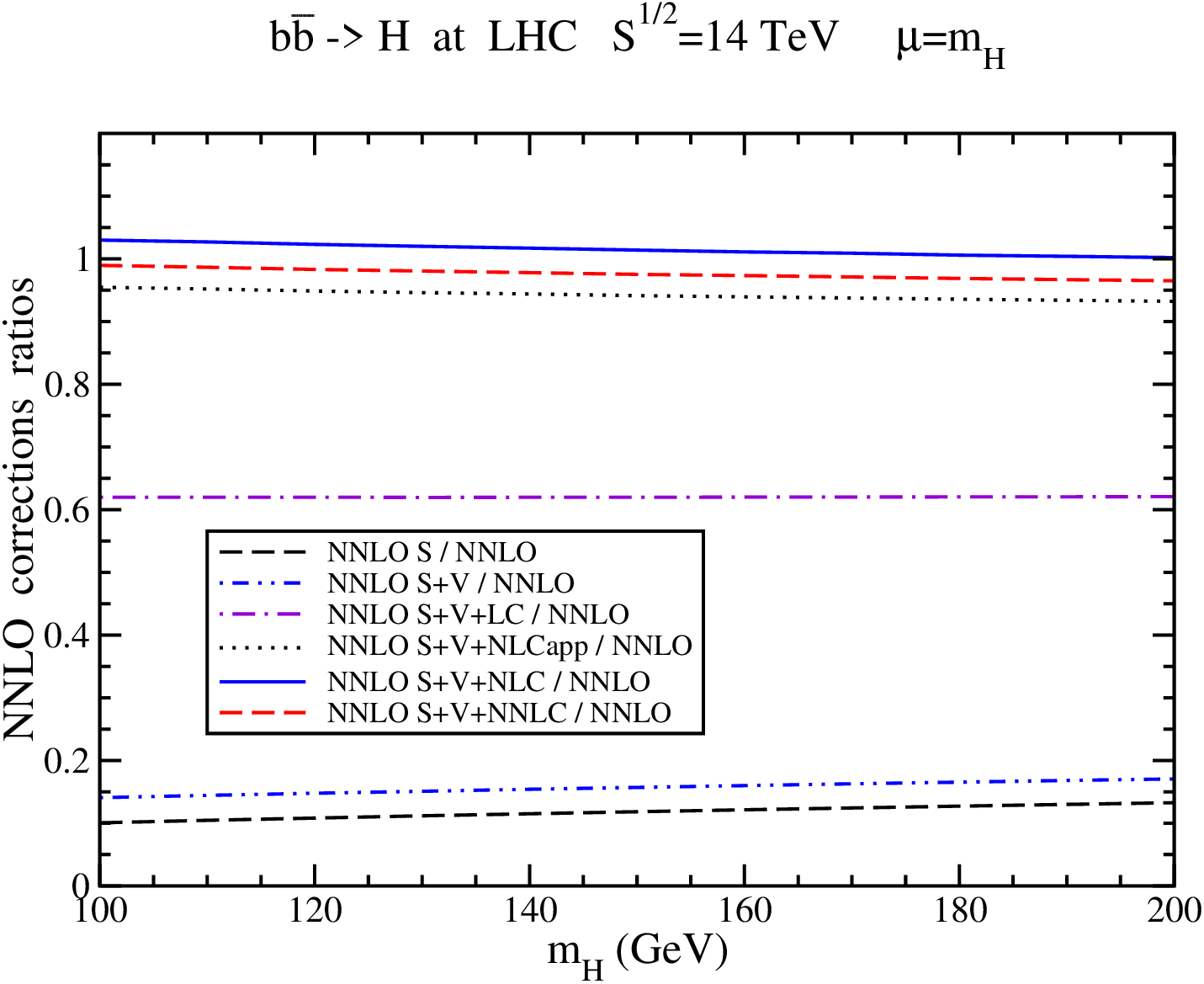}
\caption{Left: The NNLO ratios for $b {\bar b} \rightarrow H$ at the Tevatron.
Here $\mu=\mu_F=\mu_R=m_H$. Right: The NNLO ratios for $b {\bar b} 
\rightarrow H$ at the LHC.}
\label{NNLObbhplot}
\end{center}
\end{figure}

In Figure 3 we investigate the contribution of various terms 
to the complete NNLO corrections for Higgs production via 
$b {\bar b} \rightarrow H$ at the Tevatron (left-hand side) 
and the LHC (right-hand side), with $\mu=m_H$, 
by plotting the corresponding ratios. 
It is important to clarify that in this figure NNLO denotes 
the ${\cal O}(\alpha_s^2)$
corrections only (i.e. without the Born term and NLO corrections).
The curve marked NNLO S / NNLO denotes the percentage contribution of the 
NNLO soft corrections to the total NNLO corrections. We see that both at the 
Tevatron and the LHC the soft contribution is rather small, in fact even 
smaller than the relative contribution at NLO. The same holds for the 
S+V contribution. Inclusion of the leading collinear logarithms improves the 
situation but is not by itself satisfactory since it only accounts for 
about 60\% of the total NNLO corrections at both the Tevatron and the LHC. 
However, including the next-to-leading collinear logarithms vastly improves 
the approximation. The effect of the NLC terms is much more significant at 
NNLO than at NLO. As we saw in Fig. 1 and 2 at NLO the difference between 
the S+V+LC and the S+V+NLC curves was of the order of 5\% at the Tevatron 
and 10\% at the LHC, so it was not overly significant. However at NNLO
the the difference between the S+V+LC and the S+V+NLC curves is around 30\% 
at the Tevatron and  40\% at the LHC and thus the NLC terms are of utmost 
importance to get a good approximation. 

We actually plot two curves including next-to-leading collinear terms in 
Fig. 3. The curve S+V+NLC that we just discussed includes the full 
next-to-leading collinear logarithms. The curve S+V+NLCapp includes the 
approximate next-to-leading collinear logarithms from the expansion of the 
resummed cross section, Eq. (\ref{NNLOcbbh}).  
As noted before in Section 3, the expansion does not derive the 
full next-to-leading collinear logarithms; however, both at the analytical 
level and now as seen at the numerical level the difference between the two 
results is relatively small. At both the Tevatron and the LHC the S+V+NLCapp 
result accounts for around 90\% of the total NNLO corrections while the 
S+V+NLC result accounts for nearly 100\% of them, which is rather remarkable.
Finally, we also plot a curve (S+V+NNLC) that in addition includes the exact 
next-to-next-to-leading collinear terms (if instead we add the 
approximate NNLC terms from Eq. (\ref{NNLOcbbh}) the resulting curve is 
practically indistinguishable because the  NNLC approximate corrections are 
numerically very close to the exact NNLC corrections). 
From the figure we see that the NNLC terms alone do not make a large 
contribution, and that the S+V+NNLC results approximate the exact NNLO 
corrections very well. 

\begin{figure}
\begin{center}
\hspace{-5mm}
\includegraphics[width=8cm]{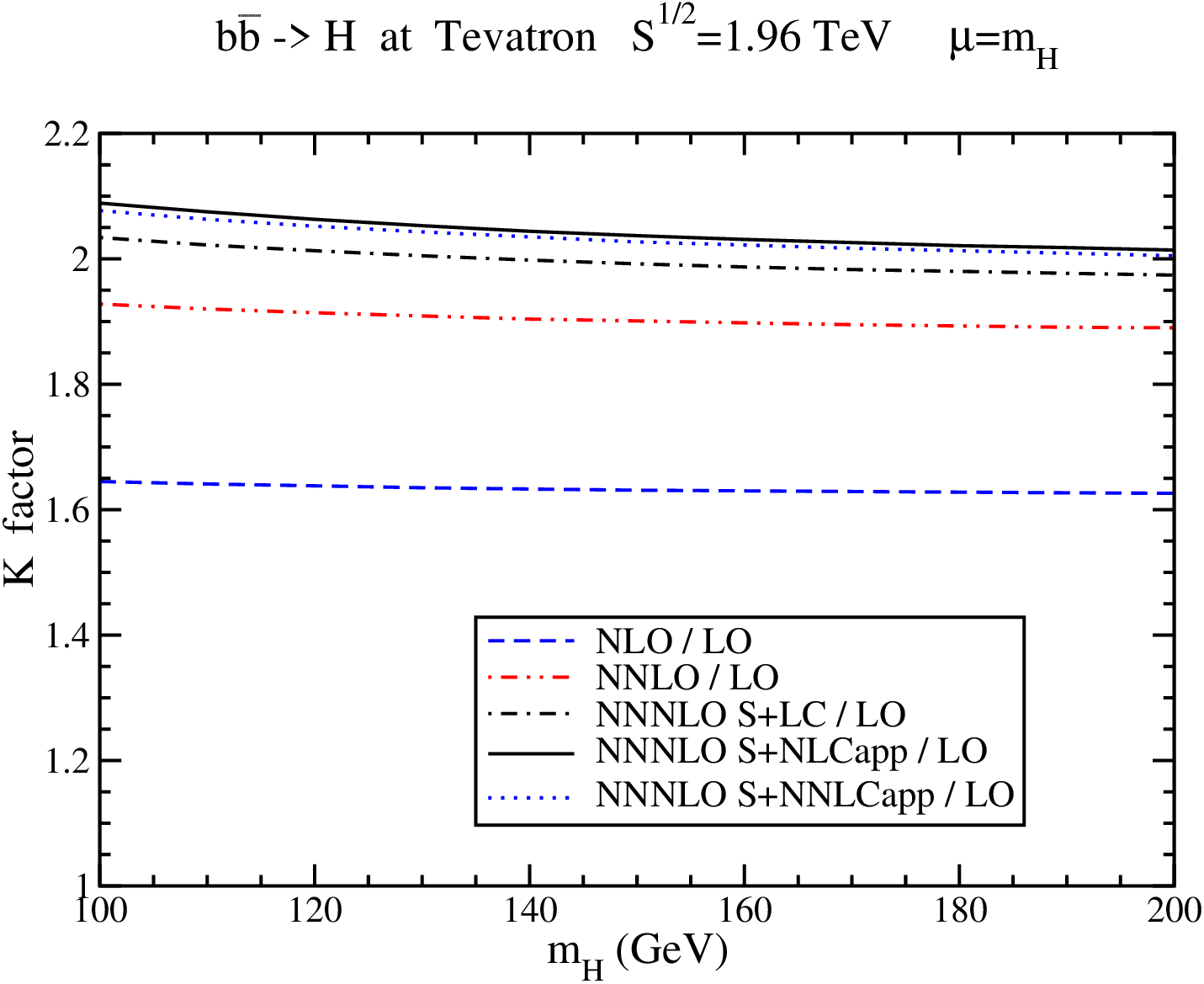}
\hspace{5mm}
\includegraphics[width=8cm]{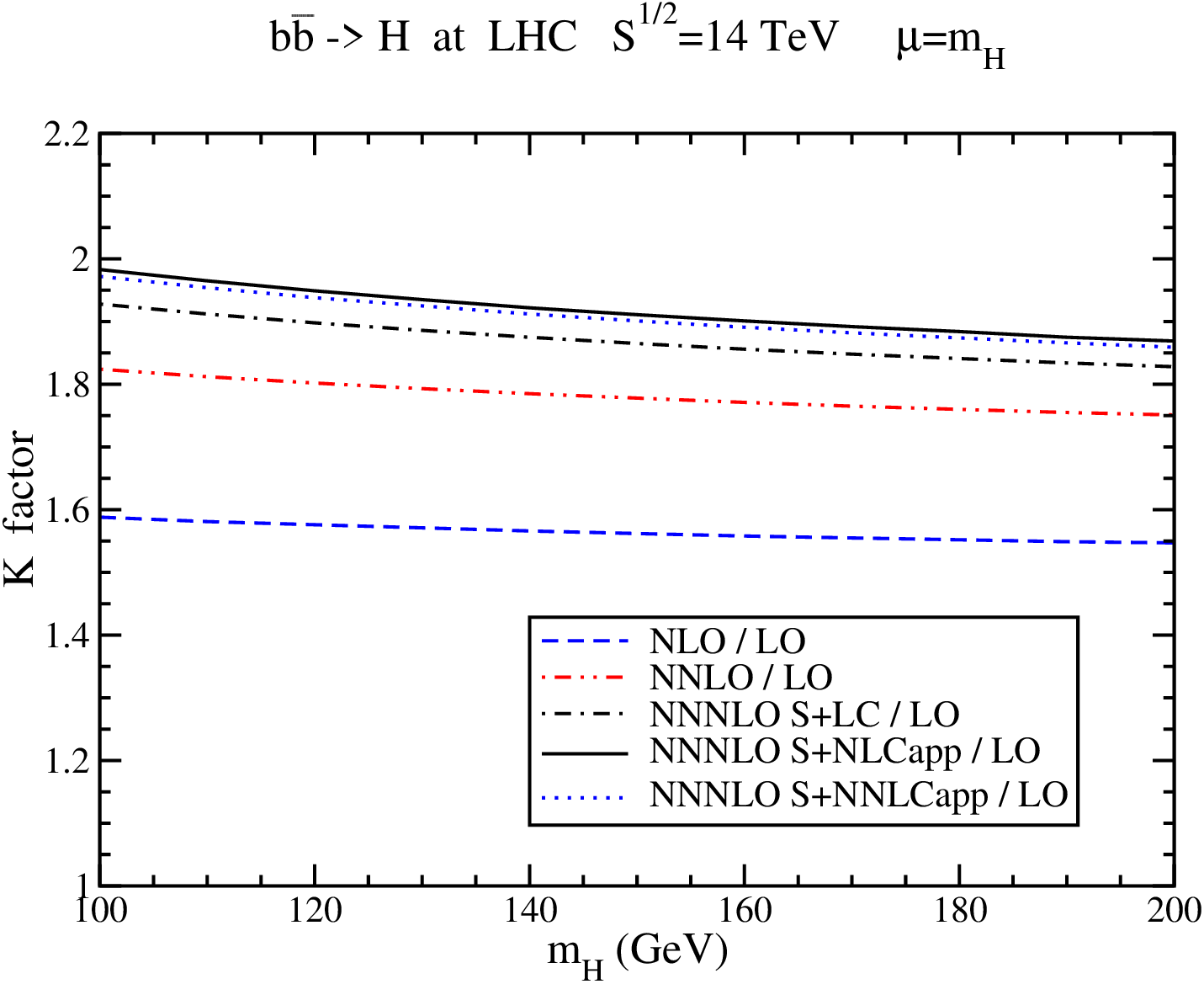}
\caption{Left: The $K$ factors for $b {\bar b} \rightarrow H$ at the Tevatron.
Here $\mu=\mu_F=\mu_R=m_H$. Right: The $K$ factors for $b {\bar b} 
\rightarrow H$ at the LHC.}
\label{Kbbhplot}
\end{center}
\end{figure}

Figure 4 shows the $K$ factors, i.e. the ratios of the cross sections 
at various orders, for Higgs production via 
$b {\bar b} \rightarrow H$ at the Tevatron (left-hand side) 
and the LHC (right-hand side), with $\mu=m_H$. 
Here and in the rest of the figures NLO cross section means 
the Born term plus the ${\cal O}(\alpha_s)$ corrections, NNLO cross section 
means the Born term plus the ${\cal O}(\alpha_s)$ and ${\cal O}(\alpha_s^2)$ 
corrections, and NNNLO cross section means the the Born term plus the 
${\cal O}(\alpha_s)$ and ${\cal O}(\alpha_s^2)$ and ${\cal O}(\alpha_s^3)$ 
corrections. We note that we have used 
the same pdf for all curves because we are interested in the relative 
size of the terms in the perturbative expansion without the 
additional complications from different pdf. As the NLO / LO 
curve shows, the complete NLO corrections increase the LO result by around 
60\% at both the Tevatron and the LHC. The NNLO / LO curve shows that inclusion
of the complete NNLO corrections futher increases the cross section by a 
substantial amount. The NNLO $K$ factor is around 1.9 at the Tevatron and 
1.8 at the LHC. Finally, we include the approximate corrections 
(soft and collinear) at NNNLO. We note that the soft corrections are 
complete and we plot one curve with the soft and leading collinear (S+LC) 
terms, another curve with the soft and approximate next-to-leading 
collinear (S+NLCapp) terms, and a third with the soft and approximate 
next-to-next-to-leading collinear (S+NNLCapp) terms.
We note that the difference between the S+LC and S+NLCapp curves is not 
very big, and between the S+NLCapp and S+NNLCapp curves it is quite small. 
Our investigation of the 
contributions of the soft and collinear terms at NLO and NNLO at both 
the Tevatron and the LHC gives us confidence that the NNNLO S+NNLCapp curves 
provide a good approximation of the complete NNNLO cross section. 
We note that the NNNLO S+NNLCapp $K$ factor is between 2.00 and 2.08 at the 
Tevatron and between 1.86 and 1.97 at the LHC for Higgs masses ranging between 
100 and 200 GeV. In both cases the NNNLO corrections provide a significant 
enhancement over the NNLO result. 

\begin{figure}
\begin{center}
\hspace{-5mm}
\includegraphics[width=8cm]{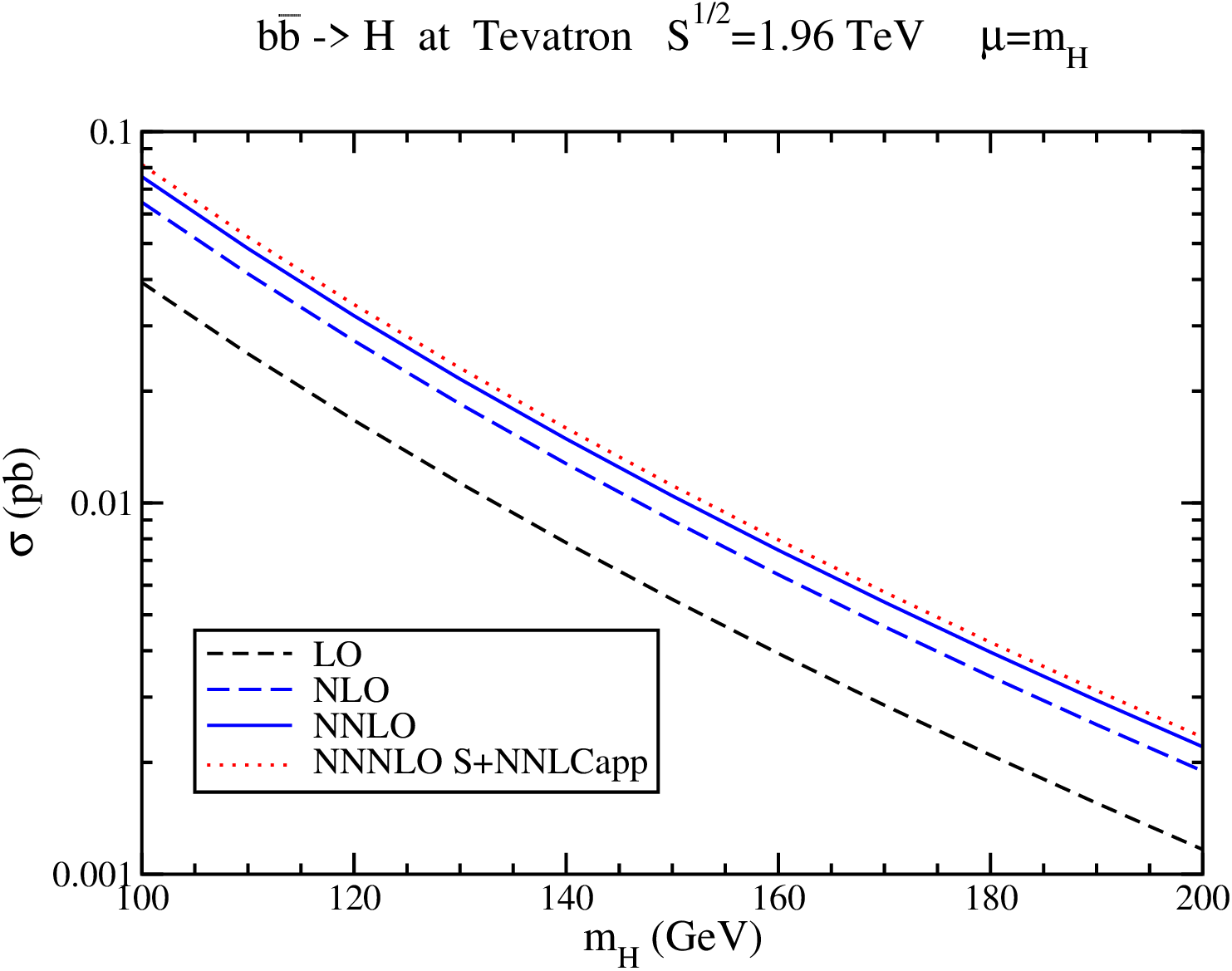}
\hspace{5mm}
\includegraphics[width=8cm]{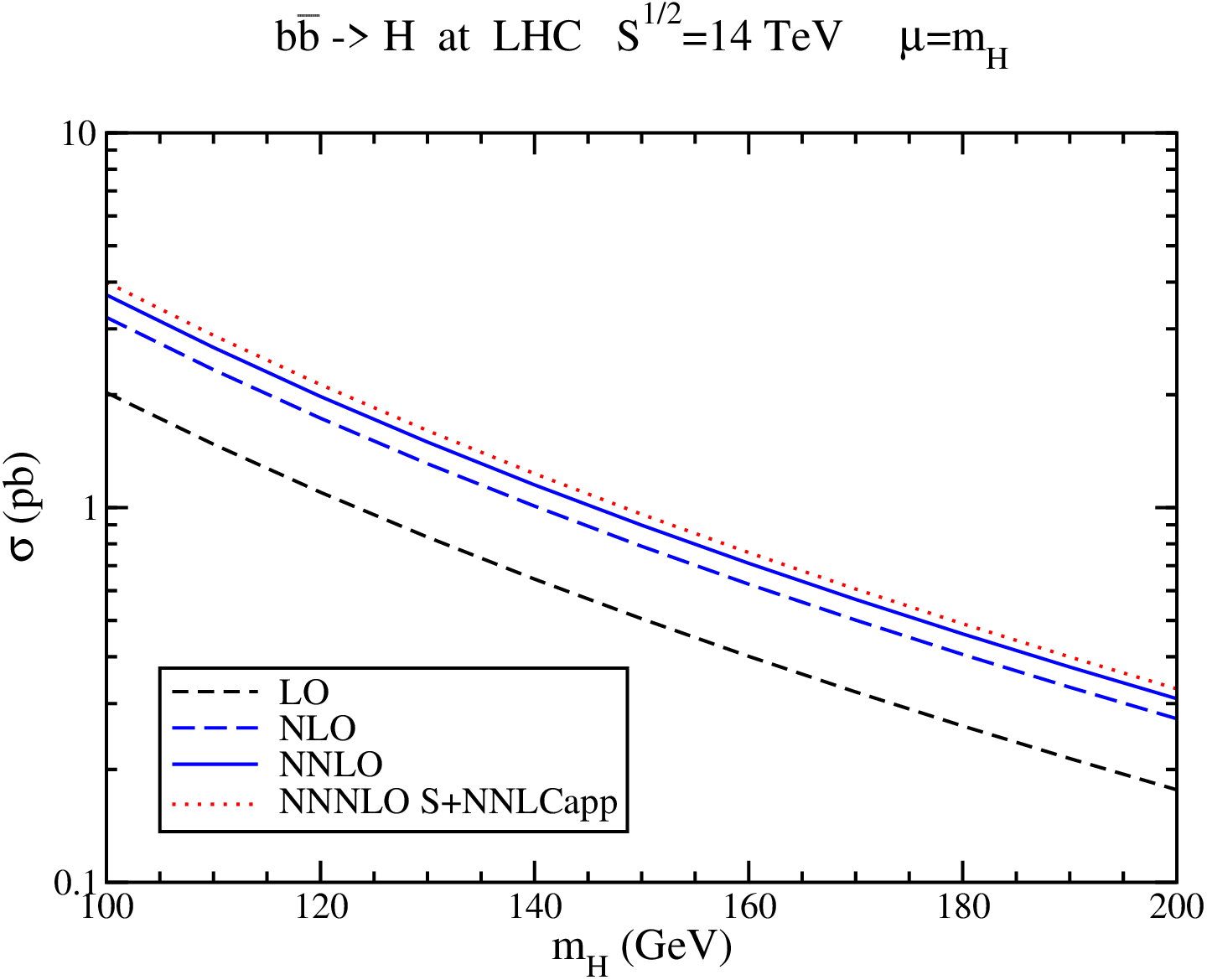}
\caption{Left: The cross section for $b {\bar b} \rightarrow H$ at the 
Tevatron with $\mu=\mu_F=\mu_R=m_H$. 
Right: The cross section for $b {\bar b} \rightarrow H$ at the LHC.}
\label{bbhmHplot}
\end{center}
\end{figure}

In Figure 5 we plot the cross sections for $b {\bar b} \rightarrow H$ at 
the Tevatron (left-hand side) and the LHC (right-hand side). We show LO, 
NLO, NNLO, and NNNLO S+NNLCapp results for $\mu=m_H$. We note that the 
NNLO SV+NLC(app) or NNLO SV+NNLC(app) results are indistinguishable from 
the exact NNLO curve on this plot. All five NNLO curves are on top of each 
other. 
This once again shows what we saw in more detail in figure 3, i.e. that 
our approximations including soft+virtual+next-to-(next-to)-leading collinear
terms are excellent.
Also, as expected from Figure 4, the NNNLO S+NLCapp curve is 
indistinguishable from NNNLO S+NNLCapp.

\begin{figure}
\begin{center}
\hspace{-5mm}
\includegraphics[width=8cm]{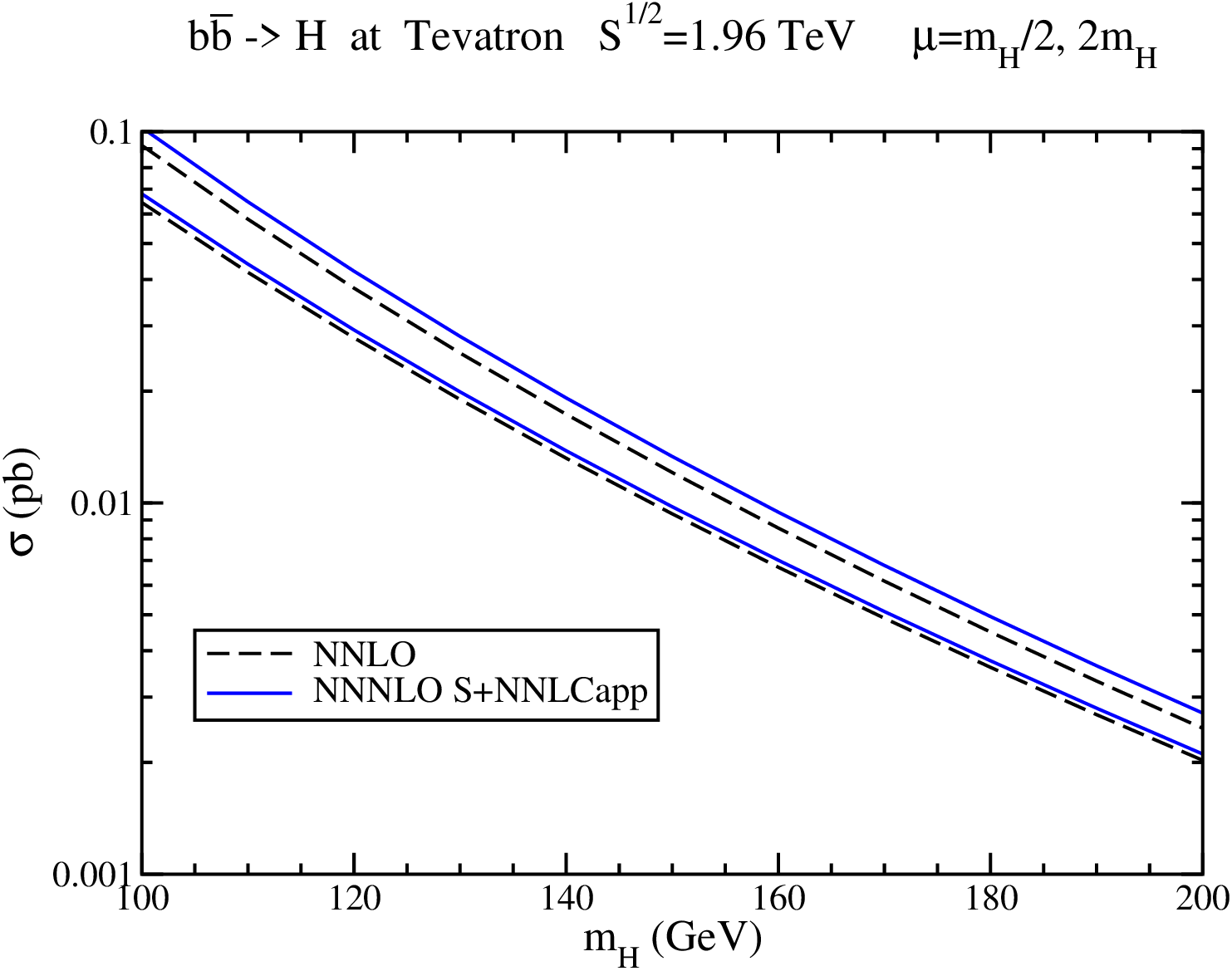}
\hspace{5mm}
\includegraphics[width=8cm]{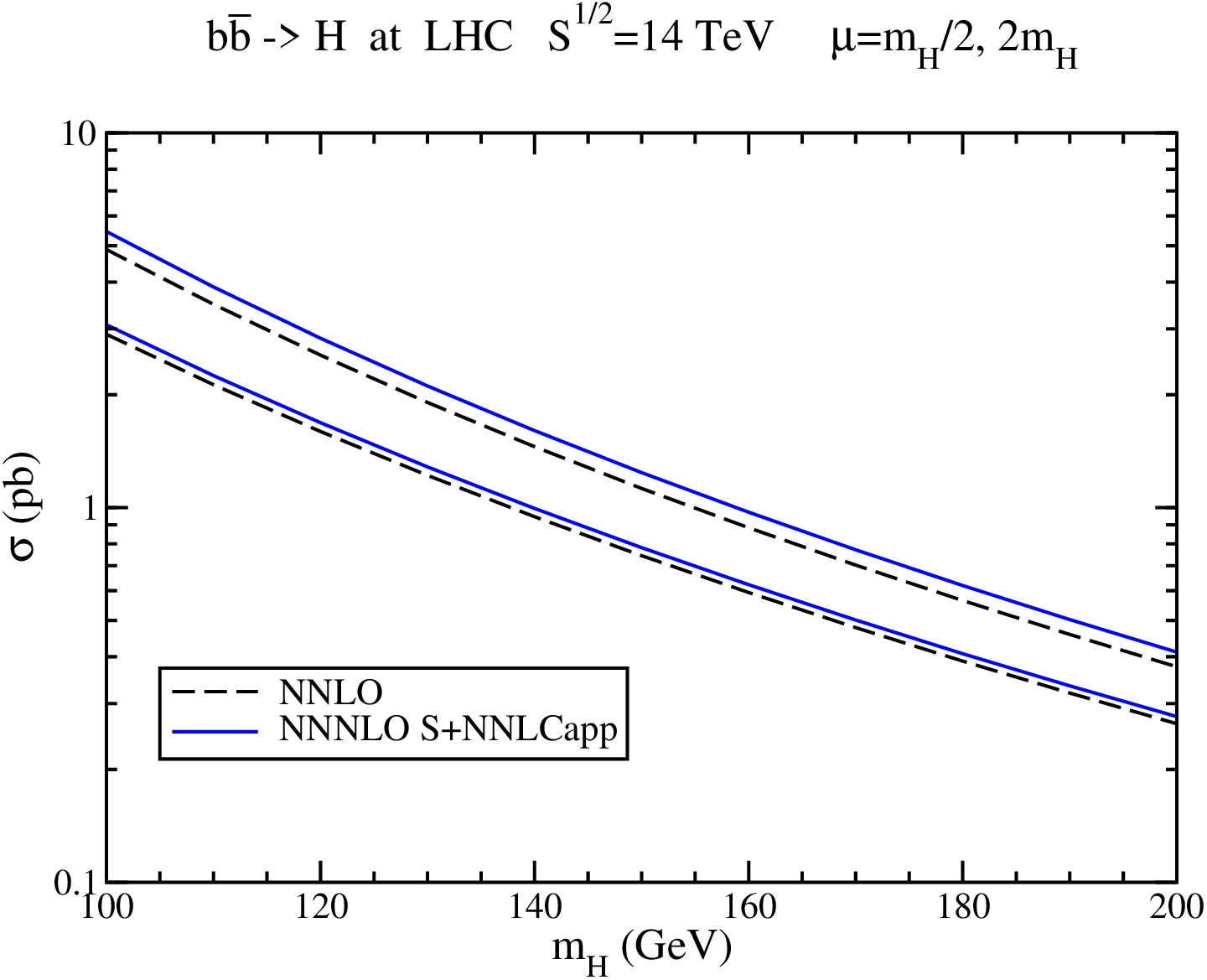}
\caption{Left: The cross section for $b {\bar b} \rightarrow H$ at the 
Tevatron with $\mu=\mu_F=\mu_R=m_H/2, 2m_H$. 
Right: The cross section for $b {\bar b} \rightarrow H$ at the LHC.}
\label{bbhmumHplot}
\end{center}
\end{figure}

In Figure 6 we plot the cross sections for $b {\bar b} \rightarrow H$ at 
the Tevatron (left-hand side) and the LHC (right-hand side) and show 
NNLO and NNNLO S+NNLCapp results for two choices of scale, $\mu=m_H/2, 2m_H$. 
We note that if we plotted instead the NNNLO S+NLCapp results they would be 
indistinguishable from the NNNLO S+NNLCapp curves. From the figure we see that 
the NNNLO S+NNLCapp result has very similar scale variation to the NNLO result.
 
Finally, we calculate the uncertainty in the cross section from the 
parton distribution functions using the sets in \cite{MRST2006}  
(pdf uncertainties for this process have also been studied in \cite{BPTY}).
We find that the uncertainty is smaller at the LHC than at the Tevatron. 
For $b {\bar b}\rightarrow H$ at the Tevatron 
we find that the pdf uncertainty varies from  3.1\% for $m_H=100$ GeV
to 5.6\% for $m_H=150$ GeV to 8.0\% for $m_H=200$ GeV.
For $b {\bar b} \rightarrow H$ at the LHC  
the pdf uncertainty varies from 2.0\% for $m_H=100$ GeV
to 1.6\% for $m_H=150$ GeV to 1.3\% for $m_H=200$ GeV.
We thus see that the pfd uncertainty is non-negligible and can be 
of the same order of magnitude as the scale uncertainty, especially for 
large Higgs masses at the Tevatron.

\mysection{Conclusions}

We have studied Higgs production at the Tevatron and the LHC  
via the channel $b {\bar b} \rightarrow H$.
We have calculated the complete soft corrections and the approximate 
next-to-next-to-leading collinear terms through NNNLO.
We have shown that the inclusion of collinear corrections is essential
in providing a good approximation to the complete cross section at NNLO.
The soft and collinear NNNLO corrections provide significant  
enhancements to the cross section and must be taken into consideration 
for an improved theoretical prediction. 
The scale dependence of the cross section and the pdf uncertainties 
were also calculated.
Analytical expressions for the soft and collinear corrections through NNNLO
were also provided for $gg \rightarrow H$ and the Drell-Yan process.

\mysection*{Acknowledgements}
 
This work was supported by the National Science Foundation under
Grant No. PHY 0555372. I would like to thank Jack Smith and Sasha Belyaev 
for useful conversations on Higgs production.
I also thank V. Ravindran for detailed comparisons of our NNNLO calculations.
 
\renewcommand{\theequation}{A.\arabic{equation}}
\setcounter{equation}{0}
\section*{Appendix A: NNNLO soft and collinear corrections for 
Higgs and Drell-Yan processes: useful formulas}

In this Appendix we first provide detailed expressions for the 
quantities that appear in the resummed cross section, Eq. (\ref{resHS}). 
We then calculate a number of integrals for the soft and collinear corrections.
Finally, we derive general formulas for the soft and collinear corrections 
for the processes under consideration in this paper.

\subsection*{A.1 Exponents in the resummed cross section}

For the quantity $A_i$ in the first exponent of Eq. (\ref{resHS}) 
we use the expansion
$A_i(\alpha_s) = A_i^{(1)} {\alpha_s/\pi}+A_i^{(2)} ({\alpha_s/\pi})^2
+A_i^{(3)}({\alpha_s/\pi})^3+\cdots$. The results differ for quarks, $i=q$, 
and gluons, $i=g$. Here 
$A_q^{(1)}=C_F=(N_c^2-1)/(2N_c)$ with $N_c=3$ the number of colors,
and $A_g^{(1)}=C_A=N_c$; 
$A_q^{(2)}=C_F K/2$ with $K= C_A(67/18-\zeta_2)-5n_f/9$ \cite{KT},
where $n_f$ is the number of light quark flavors ($n_f=5$ in our numerical 
results), and $A_g^{(2)}=C_A K/2$; 
and $A_i^{(3)}$ is given by \cite{MVV}
\beqa
A_q^{(3)}&=&C_F\left[C_A^2\left(\frac{245}{96}-\frac{67}{36}\zeta_2
+\frac{11}{24}\zeta_3+\frac{11}{20}\zeta_2^2\right)
+C_F n_f\left(-\frac{55}{96}+\frac{\zeta_3}{2}\right) \right.
\nonumber \\ && \left.
{}+C_A n_f \left(-\frac{209}{432}+\frac{5}{18}\zeta_2
-\frac{7}{12} \zeta_3\right)-\frac{n_f^2}{108}\right] \, ,
\eeqa
\beq
A_g^{(3)}=\frac{C_A}{C_F} A_q^{(3)} \, .
\eeq
Here and below $\zeta_2=\pi^2/6$, $\zeta_3=1.2020569\cdots$,  
$\zeta_4=\pi^4/90$, $\zeta_5=1.0369278\cdots$.  

Also ${\nu}_i=(\alpha_s/\pi){\nu}_i^{(1)}+(\alpha_s/\pi)^2 {\nu}_i^{(2)}
+(\alpha_s/\pi)^3 {\nu}_i^{(3)}+\cdots$, with ${\nu}_q^{(1)}=C_F$ and
${\nu}_g^{(1)}=C_A$.
It is convenient to lump together $\nu_i$ with $\Gamma_S$ as we discuss 
below.

In the second exponent of Eq. (\ref{resHS}), 
$\gamma_{i/i}$ is the moment-space 
anomalous dimension of the ${\overline {\rm MS}}$ density $\phi_{i/i}$
\cite{GALY,GFP}.
We write
\beq
\gamma_{i/i}=\gamma_i^{(N)} \ln N + \gamma_i
\eeq
with 
\beqa
\gamma_q^{(N)}&=&-\frac{\alpha_s}{\pi} C_F \ln N
-\left(\frac{\alpha_s}{\pi}\right)^2 C_F \frac{K}{2} \ln N+ \cdots
\nonumber \\ 
\gamma_g^{(N)}&=&-\frac{\alpha_s}{\pi} C_A \ln N
-\left(\frac{\alpha_s}{\pi}\right)^2 C_A \frac{K}{2} \ln N +\cdots \, ,
\eeqa
and the parton anomalous dimensions
\beq
\gamma_i=(\alpha_s/\pi) \gamma_i^{(1)}
+(\alpha_s/\pi)^2 \gamma_i^{(2)} + \cdots
\eeq
with  $\gamma_q^{(1)}=3C_F/4$, $\gamma_g^{(1)}=\beta_0/4$,
\beq
\gamma_q^{(2)}=C_F^2\left(\frac{3}{32}-\frac{3}{4}\zeta_2
+\frac{3}{2}\zeta_3\right)
+C_F C_A\left(\frac{17}{96}+\frac{11}{12}\zeta_2-\frac{3}{4}\zeta_3\right)
+n_f C_F \left(-\frac{1}{48}-\frac{\zeta_2}{6}\right)\, ,
\eeq
and
\beq
\gamma_g^{(2)}=C_A^2\left(\frac{2}{3}+\frac{3}{4}\zeta_3\right)
-n_f\left(\frac{C_F}{8}+\frac{C_A}{6}\right) \, .
\eeq

The $\beta$ function in the third exponent of Eq. (\ref{resHS}) is given by
\beq
\beta(\alpha_s) \equiv \frac{1}{2\alpha_s}\frac{d\alpha_s}{d\ln \mu}
=\mu \, d \ln g/d \mu
=-\beta_0 \alpha_s/(4 \pi)-\beta_1 \alpha_s^2/(4 \pi)^2+\cdots \, ,
\eeq
where $g^2=4 \pi \alpha_s$,
with $\beta_0=(11C_A-2n_f)/3$ and $\beta_1=34 C_A^2/3-2n_f(C_F+5C_A/3)$ 
\cite{}.
Note that
\beqa
\alpha_s(\mu)&=&\alpha_s(\mu_R)\left[1-\frac{\beta_0}{4\pi}\alpha_s(\mu_R)
\ln\left(\frac{{\mu}^2}{\mu_R^2}\right)
+\frac{\beta_0^2}{16\pi^2}\alpha_s^2(\mu_R)
\ln^2\left(\frac{{\mu}^2}{\mu_R^2}\right)
-\frac{\beta_1}{16\pi^2}\alpha_s^2(\mu_R)
\ln\left(\frac{{\mu}^2}{\mu_R^2}\right) \right.
\nonumber \\ && \hspace{20mm} \left.
+\cdots\right] \, .
\eeqa
Also $d_{\alpha_s}$ is 0 for $b {\bar b} \rightarrow H$ and 
$q {\bar q} \rightarrow V$ and 2 for $gg \rightarrow H$. 

The hard and soft functions can be expanded as 
$H=\alpha_s^{d_{\alpha_s}}H^{(0)}+(\alpha_s^{d_{\alpha_s}+1}/\pi)H^{(1)}
+(\alpha_s^{d_{\alpha_s}+2}/\pi^2)H^{(2)}$\linebreak
$+(\alpha_s^{d_{\alpha_s}+3}/\pi^3)H^{(3)}
+\cdots$ and
$S=S^{(0)}+(\alpha_s/\pi)S^{(1)}+(\alpha_s/\pi)^2 S^{(2)}
+(\alpha_s/\pi)^3 S^{(3)}+\cdots$.
At lowest order, the Born cross section for the partonic process is
$\sigma^B=\alpha_s^{d_{\alpha_s}} H^{(0)}S^{(0)}$.

The soft anomalous dimension is denoted by $\Gamma_S$ and can be 
expanded as $\Gamma_S=(\alpha_s/\pi)\Gamma_S^{(1)}$ \linebreak
$+(\alpha_s/\pi)^2\Gamma_S^{(2)}+(\alpha_s/\pi)^3\Gamma_S^{(3)}+\cdots$.
For the processes $b{\bar b}\rightarrow H$ and $q{\bar q}\rightarrow V$,
$\Gamma^{(1)}_{S\, q {\bar q}}=C_F$. 
For the process $gg\rightarrow H$, $\Gamma^{(1)}_{S\, gg}=C_A$.
If we define $G^{(n)}=\Gamma_S^{(n)}-\nu_i^{(n)}$ we can get 
explicit expressions through $G^{(3)}$ \cite{MVNNNLO,LM,VR2}. We have 
$G_{q {\bar q}}^{(1)}=G_{gg}^{(1)}=0$, 
\beq
G^{(2)}_{q \bar q}=C_F C_A \left(-\frac{101}{54}+\frac{11}{6} \zeta_2
+\frac{7}{4}\zeta_3\right)
+\frac{1}{3} C_F n_f \left(\frac{7}{9}-\zeta_2\right) \, ,
\eeq
\beq
G^{(2)}_{gg}=\frac{C_A}{C_F}G^{(2)}_{q \bar q} \, , 
\eeq
\beqa
G^{(3)}_{q \bar q}&=&C_F C_A^2 \left(-\frac{297029}{46656}
+\frac{6139}{648} \zeta_2+\frac{2509}{216} \zeta_3-3 \zeta_5
-\frac{187}{120} \zeta_2^2 -\frac{11}{12} \zeta_2 \zeta_3\right)
\nonumber \\ && 
{}+C_F C_A n_f \left(\frac{31313}{23328}-\frac{1837}{648}\zeta_2
-\frac{155}{72}\zeta_3+\frac{23}{60}\zeta_2^2\right)
\nonumber \\ && 
{}+C_F^2 n_f \left(\frac{1711}{1728}-\frac{\zeta_2}{4}-\frac{19}{36}\zeta_3
-\frac{\zeta_2^2}{10}\right)
+C_F n_f^2 \left(-\frac{29}{729}+\frac{5}{27}\zeta_2
+\frac{5}{54}\zeta_3 \right) \, ,
\eeqa
\beq
G^{(3)}_{gg}=\frac{C_A}{C_F}G^{(3)}_{q \bar q} \, .
\eeq

\subsection*{A.2 Useful integrals}

In the resummed cross section we encounter certain integrals 
that need to be evaluated.
They are of the form 
\beq
I_n(N)=\int_0^1 dz \, z^{N-1} \left[\frac{\ln^n(1-z)}{1-z}\right]_+ \, .
\eeq
The expressions for  $I_n$ have been presented up to $n=7$ in
\cite{NKtop} and including $1/N$ terms up to $n=3$ in \cite{KLS}.  
Here we extend these results by including $1/N$ terms up to $n=5$, 
which are needed for NNNLO expansions:
\beqa
I_0(N)&=& -\ln{\tilde N}+\frac{1}{2N}
\nonumber \\
I_1(N)&=&\frac{1}{2} \ln^2{\tilde N}+\frac{\zeta_2}{2}
-(\ln{\tilde N}+1) \frac{1}{2N}
\nonumber \\
I_2(N)&=&-\frac{1}{3} \ln^3{\tilde N}-\zeta_2 \ln{\tilde N}
-\frac{2}{3}\zeta_3+\left(\frac{1}{2} \ln^2{\tilde N}+\frac{\zeta_2}{2}
+\ln{\tilde N}\right)\frac{1}{N}
\nonumber \\
I_3(N)&=& \frac{1}{4} \ln^4{\tilde N}+\frac{3}{2} \zeta_2 \ln^2{\tilde N}
+2 \zeta_3 \ln{\tilde N} +\frac{3}{2} \zeta_4
+\frac{3}{4} \zeta_2^2
\nonumber \\ &&
{}+\left(-\frac{1}{2}\ln^3{\tilde N}
-\frac{3}{2}\zeta_2 \ln{\tilde N}-\zeta_3-\frac{3}{2} \ln^2{\tilde N}
-\frac{3}{2}\zeta_2\right)\frac{1}{N}
\nonumber \\
I_4(N)&=& -\frac{1}{5} \ln^5{\tilde N} -2 \zeta_2 \ln^3{\tilde N}
-4 \zeta_3 \ln^2{\tilde N} -3 (\zeta_2^2+2 \zeta_4) \ln{\tilde N} 
-4\left(\zeta_2 \zeta_3+\frac{6}{5} \zeta_5\right)
\nonumber \\ &&
{}+\left(\frac{1}{2} \ln^4{\tilde N}+3 \zeta_2 \ln^2{\tilde N}
+4 \zeta_3 \ln{\tilde N} +\frac{3}{2} \zeta_2^2 +3 \zeta_4
+2 \ln^3{\tilde N}+6 \zeta_2 \ln{\tilde N} +4 \zeta_3\right) \frac{1}{N}
\nonumber \\
I_5(N)&=& \frac{1}{6} \ln^6{\tilde N}+\frac{5}{2} \zeta_2 \ln^4{\tilde N}
+\frac{20}{3} \zeta_3 \ln^3{\tilde N}
+\frac{15}{2}(\zeta_2^2+2\zeta_4) \ln^2{\tilde N}
\nonumber \\ &&
{}+4(5\zeta_2 \zeta_3+6\zeta_5) \ln{\tilde N}
+5\left(\frac{\zeta_2^3}{2}+\frac{4}{3}\zeta_3^2+3\zeta_2\zeta_4
+4\zeta_6\right)
\nonumber \\ &&
{}+\left(-\frac{1}{2} \ln^5{\tilde N} -5 \zeta_2 \ln^3{\tilde N}
-10 \zeta_3 \ln^2{\tilde N} -\frac{15}{2} (\zeta_2^2+2 \zeta_4) \ln{\tilde N}
-10\left(\zeta_2 \zeta_3+\frac{6}{5} \zeta_5 \right) \right.
\nonumber \\ && \quad \left.
{}-\frac{5}{2} \ln^4{\tilde N}-15 \zeta_2 \ln^2{\tilde N}
-20 \zeta_3 \ln{\tilde N} -\frac{15}{2} \zeta_2^2 -15 \zeta_4\right)
\frac{1}{N} \, .
\eeqa

We also encounter integrals of the form 
\beq
J_n(N)=\int_0^1 dz \, z^{N-1} \ln^n(1-z) \, .
\eeq
The expressions for  $J_n$ have been presented up to $n=3$ in
\cite{KLS}. 
Here we extend these results up to $n=5$, 
which are needed for NNNLO expansions:
\beqa
J_0(N)&=& \frac{1}{N}
\quad \quad \quad
J_1(N)= -\frac{\ln{\tilde N}}{N}
\quad \quad \quad
J_2(N)= \frac{\ln^2{\tilde N}}{N}+\frac{\zeta_2}{N}
\nonumber \\
J_3(N)&=& -\frac{\ln^3{\tilde N}}{N}-3\zeta_2\frac{\ln{\tilde N}}{N}
-2\frac{\zeta_3}{N}
\nonumber \\
J_4(N)&=& \frac{\ln^4{\tilde N}}{N}+6\zeta_2\frac{\ln^2{\tilde N}}{N}
+8\zeta_3\frac{\ln{\tilde N}}{N}
+(3\zeta_2^2+6\zeta_4)\frac{1}{N}
\nonumber \\
J_5(N)&=& -\frac{\ln^5{\tilde N}}{N}-10\zeta_2\frac{\ln^3{\tilde N}}{N}
-20\zeta_3\frac{\ln^2{\tilde N}}{N}
-15(\zeta_2^2+2\zeta_4)\frac{\ln{\tilde N}}{N}
-20\left(\zeta_2 \zeta_3+\frac{6}{5}\zeta_5\right)\frac{1}{N} \, .
\nonumber \\
\eeqa

\subsection*{A.3 General formulas for soft and collinear corrections}

Below, we write general expressions for the soft and collinear corrections 
at NLO, NNLO, and NNNLO for the processes under study. 
Detailed results are provided in Section 3 for 
$b {\bar b} \rightarrow H$, in Appendix B for the Drell-Yan process, and 
in Appendix C for $gg \rightarrow H$.
We note that for processes involving a quark running mass, such as 
$b {\bar b} \rightarrow H$, the expressions below are valid for the 
running mass at scale $M$. If the mass is evaluated at scale $\mu_R$ 
(as in Section 3) we have to include additional terms as discussed at the end 
of this appendix.
  
The NLO soft and virtual corrections are given by
\beq
{\hat{\sigma}}^{(1)\, SV} = F^B \frac{\alpha_s(\mu_R^2)}{\pi}
\left\{c_3\, {\cal D}_1(z) + c_2\,  {\cal D}_0(z) 
+c_1\,  \delta(1-z)\right\}
\label{NLO}
\eeq
where $F^B$ is the Born term and
\beq
D_l(z)=\left[\frac{\ln^l(1-z)}{1-z}\right]_+
\eeq
with $z=M^2/s$.
If we define $C_i$ to denote $C_F$ for quarks and $C_A$ for gluons then 
the coefficients are given by 
\beq
c_3=4 C_i 
\eeq
\beq
c_2=-2 C_i \ln\left(\frac{\mu_F^2}{M^2}\right)
\eeq
\beq
c_1=T_1+c_1^{\mu}
\eeq
with
\beq
c_1^{\mu}=-2 \gamma_i^{(1)} \ln\left(\frac{\mu_F^2}{M^2}\right)
+d_{\alpha_s} \frac{\beta_0}{4} \ln\left(\frac{\mu_R^2}{M^2}\right) \, ,
\eeq
and $T_1$ the scale-independent virtual corrections for each process.
We note that $c_2$ only involves the scale dependence. This 
simplifies the structure of the results at higher orders.

The NLO collinear corrections are given by                                     
\beq               
{\hat{\sigma}}^{(1)\, C} = F^B \frac{\alpha_s(\mu_R^2)}{\pi}
\left\{c_3^c\, \ln(1-z) + c_2^c \right\}
\label{NLOc} 
\eeq               
where $c_3^c=-4C_i \kappa_i$ and $c_2^c=2C_i \kappa_i \ln(\mu_F^2/M^2)+2C_i$,
where $\kappa_q=1$ and $\kappa_g=2$.

The NNLO soft and virtual corrections are given by
\beqa
{\hat{\sigma}}^{(2)\, SV}&=&F^B \frac{\alpha_s^2(\mu_R^2)}{\pi^2} 
\left\{\frac{1}{2} c_3^2 \, {\cal D}_3(z)
+\left[\frac{3}{2} c_3 \, c_2 - \frac{\beta_0}{4} c_3 \right] \, 
{\cal D}_2(z) \right.
\nonumber \\ && 
{}+\left[c_3 \, c_1 +c_2^2
-\zeta_2 \, c_3^2+\frac{\beta_0}{4} c_3
\ln\left(\frac{\mu_R^2}{M^2}\right) +\frac{c_3}{2}K\right] \,
{\cal D}_1(z)
\nonumber \\ && 
{}+\left[c_2 c_1-\zeta_2 \, c_3 \, c_2
+\zeta_3 \, c_3^2 +2 G^{(2)}_{ii}
+\frac{\beta_0}{4} c_2 \ln\left(\frac{\mu_R^2}{M^2}\right) \right.
\nonumber \\ && \quad \left. 
{}+C_i\frac{\beta_0}{4} \ln^2\left(\frac{\mu_F^2}{M^2}\right)
-C_i K \ln\left(\frac{\mu_F^2}{M^2}\right) \right] \, 
{\cal D}_0(z) 
\nonumber \\ && 
{}+\left[V^{(2)}+\frac{1}{2}(c_1^2-T_1^2)
-\frac{\zeta_2}{2}c_2^2+\zeta_3 c_3 c_2
+\frac{\beta_0}{4}c_1 \ln\left(\frac{\mu_R^2}{M^2}\right)
-2{\gamma}_i^{(2)} \ln\left(\frac{\mu_F^2}{M^2}\right)\right.
\nonumber \\ && \quad \left. \left. 
{}+\frac{\beta_0}{4} {\gamma}_i^{(1)} \ln^2\left(\frac{\mu_F^2}{M^2}\right)
-\frac{d_{\alpha_s}}{32} \beta_0^2 \ln^2\left(\frac{\mu_R^2}{M^2}\right)
+\frac{d_{\alpha_s}}{16} \beta_1 \ln\left(\frac{\mu_R^2}{M^2}\right)\right] 
\delta(1-z)\right\} 
\label{NNLO}
\eeqa
where $V^{(2)}$ denotes the two-loop virtual scale-independent 
terms for each process, which cannot be derived from resummation.

The NNLO (approximate) collinear corrections are given by
\beqa               
{\hat{\sigma}}^{(2)\, C} &=& F^B \frac{\alpha_s^2(\mu_R^2)}{\pi^2}
\left\{\frac{1}{2} c_3 c_3^c\, \ln^3(1-z) 
+\left[\frac{1}{2} c_3 c_2^c +c_2 c_3^c+\frac{1}{2}c_3^2
-\frac{\beta_0}{4}c_3^c\right] \ln^2(1-z) \right.
\nonumber \\ && 
{}+\left[c_1 c_3^c+c_2 c_2^c-\zeta_2 c_3 c_3^c+c_3 c_2
+\frac{\beta_0}{4} c_3^c \ln\left(\frac{\mu_R^2}{M^2}\right)
+\frac{1}{2}c_3^c K-\frac{\beta_0}{2}T_2^c\right] \ln(1-z)
\nonumber \\ && \left. 
{}+{\rm const.} \right\}
\label{NNLOc} 
\eeqa               
where $T_2^c=2C_i$ denotes the scale-independent part of $c_2^c$.
Note that only the leading terms are complete. However, the 
$\ln^2(1-z)$ and $\ln(1-z)$ are nearly complete as discussed elsewhere in the
paper. We do not calculate the constants other than the scale-dependent terms
as described elsewhere in the paper.

The NNNLO soft-gluon corrections are given by 
\beqa
{\hat{\sigma}}^{(3)\, S}&=&
F^B \frac{\alpha_s^3(\mu_R^2)}{\pi^3} \;
\left \{\frac{1}{8} c_3^3 \; {\cal D}_5(z)
+\frac{5}{24}\left(3 c_3^2 c_2 -\beta_0 c_3^2 \right) \;
{\cal D}_4(z) \right.
\nonumber \\ && \hspace{-15mm}
{}+\left[c_3 c_2^2 +\frac{1}{2} c_3^2 c_1
-\zeta_2 c_3^3 -\frac{\beta_0}{3} c_3 c_2 +\frac{\beta_0^2}{12} c_3 
+\frac{K}{2} c_3^2 +\frac{\beta_0}{4} c_3^2 
\ln\left(\frac{\mu_R^2}{M^2}\right)\right] \; {\cal D}_3(z)
\nonumber \\ && \hspace{-15mm}
{}+\left[\frac{3}{2} c_3 c_2 \, c_1 +\frac{1}{2} c_2^3
-3 \zeta_2  c_3^2 c_2 +\frac{5}{2} \zeta_3 c_3^3
-\frac{\beta_0}{4} c_3 c_1+\frac{3}{4}\beta_0 \zeta_2 c_3^2 
+\frac{3}{4} K c_3 c_2 -\frac{\beta_0}{4} K c_3 -\frac{\beta_1}{4} C_i
\right.
\nonumber \\ && \hspace{-10mm} \left.
{}+3 c_3 G^{(2)}_{ii} +\frac{3}{8}C_i \beta_0 c_3 
\ln^2\left(\frac{\mu_F^2}{M^2}\right)-\frac{3}{2} C_i K c_3  
\ln\left(\frac{\mu_F^2}{M^2}\right)
+\left(\frac{3}{4} \beta_0 c_3 c_2 -\frac{\beta_0^2}{8} c_3\right)
\ln\left(\frac{\mu_R^2}{M^2}\right) \right]\; {\cal D}_2(z)
\nonumber \\ &&  \hspace{-15mm}
{}+\left[\frac{c_3}{2}(c_1^2-T_1^2)+c_2^2 c_1
-\zeta_2 c_3^2 c_1-\frac{5}{2} \zeta_2 c_3 c_2^2
+5 \zeta_3 c_3^2 c_2 -3 \zeta_4 c_3^3 +\zeta_2^2  c_3^3  
+\frac{\beta_0}{2} \zeta_2 c_3 c_2-\frac{3}{2} \beta_0 \zeta_3 c_3^2
\right.
\nonumber \\ && \hspace{-10mm} 
{}+\frac{K}{2} c_3 c_1+C_i \zeta_2 K c_3-\frac{5}{4} \zeta_2 K c_3^2
+(4c_2-2 \beta_0) G^{(2)}_{ii}+c_3 V^{(2)}+4 A_i^{(3)}
+\frac{\beta_0}{4}\left(c_3 \gamma_i^{(1)}+2 C_i c_2\right) 
\ln^2\left(\frac{\mu_F^2}{M^2}\right)
\nonumber \\ && \hspace{-10mm} 
{}-2 \left(c_3\gamma_i^{(2)}+C_i K c_2\right) 
\ln\left(\frac{\mu_F^2}{M^2}\right)
-\frac{\beta_0^2}{2} C_i \ln\left(\frac{\mu_F^2}{M^2}\right) 
\ln\left(\frac{\mu_R^2}{M^2}\right) 
+\frac{\beta_0^2}{32} \left(8C_i -c_3 d_{\alpha_s}\right) 
\ln^2\left(\frac{\mu_R^2}{M^2}\right)  
\nonumber \\ && \hspace{-10mm} \left.
{}+\left( \frac{\beta_0}{2} (c_3 c_1+c_2^2-\zeta_2 c_3^2+2C_i K)
-\frac{\beta_0^2}{4} c_2 + \frac{\beta_1}{16} 
(4 C_i+c_3 d_{\alpha_s})\right)
\ln\left(\frac{\mu_R^2}{M^2}\right)  \right] \; {\cal D}_1(z)
\nonumber \\ && \hspace{-15mm}
{}+\left[\frac{c_2}{2}(c_1^2-T_1^2)
-\zeta_2 c_3 c_2 c_1-\frac{\zeta_2}{2}c_2^3
+\zeta_3 c_3^2 c_1+2 \zeta_3 c_3 c_2^2
-3 \zeta_4 c_3^2 c_2 +3\zeta_5 c_3^3 +\zeta_2^2 c_3^2 c_2
-2 \zeta_2 \zeta_3 c_3^3+C_i \zeta_2 K c_2  \right.
\nonumber \\ && \hspace{-10mm}
{}-\frac{3}{4} \zeta_2 K c_3 c_2
+\zeta_3 K c_3^2-\frac{\beta_0}{2} \zeta_3 c_3 c_2 
+\frac{5}{4} \beta_0 \zeta_4 c_3^2-\frac{\beta_0}{4} c_3^2 \zeta_2^2
+2(c_1-\zeta_2 c_3) G^{(2)}_{ii}+c_2 V^{(2)}+ 2 G^{(3)}_{ii}
\nonumber \\ && \hspace{-10mm}
{}-\frac{\beta_0^2}{24}C_i \ln^3\left(\frac{\mu_F^2}{M^2}\right)
+\left(\frac{\beta_0}{4} (C_i c_1-C_i \zeta_2 c_3+C_i K+c_2\gamma_i^{(1)})
+\frac{\beta_1}{16} C_i\right) \ln^2\left(\frac{\mu_F^2}{M^2}\right)
\nonumber \\ && \hspace{-10mm}
{}+\frac{\beta_0^2}{8} C_i \ln^2\left(\frac{\mu_F^2}{M^2}\right)
\ln\left(\frac{\mu_R^2}{M^2}\right)
-\frac{\beta_0^2}{8} C_i \ln\left(\frac{\mu_F^2}{M^2}\right)
\ln^2\left(\frac{\mu_R^2}{M^2}\right)
-\frac{d_{\alpha_s}}{32}c_2 \beta_0^2 \ln^2\left(\frac{\mu_R^2}{M^2}\right)
\nonumber \\ && \hspace{-10mm}
{}-\frac{C_i}{8}(4\beta_0 K+\beta_1)\ln\left(\frac{\mu_F^2}{M^2}\right)
\ln\left(\frac{\mu_R^2}{M^2}\right)
+\left(-C_i K c_1+C_i \zeta_2 K c_3-2 c_2\gamma_i^{(2)}-2A_i^{(3)}\right)
\ln\left(\frac{\mu_F^2}{M^2}\right)
\nonumber \\ && \hspace{-10mm} \left. \left.
{}+\left(\frac{\beta_0}{2}(c_2 c_1-\zeta_2 c_3 c_2
+\zeta_3 c_3^2+2G^{(2)}_{ii})
+\frac{d_{\alpha_s}}{16}c_2 \beta_1\right) \ln\left(\frac{\mu_R^2}{M^2}\right)
\right] \; {\cal D}_0(z) \right\} \, .
\label{NNNLO}
\eeqa

The NNNLO (approximate) collinear corrections are given by
\beqa               
{\hat{\sigma}}^{(3)\, C} &=& F^B \frac{\alpha_s^3(\mu_R^2)}{\pi^3}
\left\{\frac{1}{8} c_3^2 c_3^c\, \ln^5(1-z) 
+\left[\frac{1}{8} c_3^2 c_2^c +\frac{1}{2}c_3 c_2 c_3^c-\frac{c_3^3}{16}
-\frac{5}{24}\beta_0 c_3 c_3^c\right] \ln^4(1-z) \right.
\nonumber \\ && 
{}+\left[\frac{1}{2}c_3 c_1 c_3^c+\frac{1}{2}c_2^2 c_3^c
+\frac{1}{2}c_3 c_2 c_2^c-\zeta_2 c_3^2 c_3^c-\frac{1}{4}c_3^2 c_2
+\frac{\beta_0^2}{12} c_3^c-\frac{\beta_0}{4}c_3^2
-\frac{\beta_0}{4}c_2 c_3^c-\frac{\beta_0}{12}c_3 c_2^c \right.
\nonumber \\ && \left. \left.
{}-\frac{\beta_0}{4}c_3 T_2^c+\frac{1}{2}c_3 c_3^c K
+\frac{\beta_0}{4}c_3 c_3^c \ln\left(\frac{\mu_R^2}{M^2}\right)
\right] \ln^3(1-z)+{\cal O}(\ln^2(1-z))\right\} \, .
\label{NNNLOc} 
\eeqa               
Note that only the leading terms are complete. However, the 
$\ln^4(1-z)$ and $\ln^3(1-z)$ are expected to be nearly complete 
as discussed elsewhere in the
paper. We do not calculate $\ln^2(1-z)$ or lower terms other than some  
scale-dependent terms as described elsewhere in the paper.

In the calculation of $b {\bar b} \rightarrow H$ we also have to consider 
the renormalization scale logarithms from the $\overline{\rm MS}$ 
bottom quark running mass, $m_b$.
Since $\mu^2 dm_b/d\mu^2=-\gamma_m m_b$ with $\gamma_m$ 
the mass anomalous dimension \cite{gammam}
\beq
\gamma_m=\frac{\alpha_s}{\pi} \frac{3}{4}C_F 
+\frac{\alpha_s^2}{\pi^2} \left(\frac{3}{32} C_F^2
+\frac{97}{96} C_F C_A -\frac{5}{48} n_f C_F\right)
+{\cal O}(\alpha_s^3)
\eeq
we have
\beqa
m_b^2(m_H^2)&=&m_b^2(\mu_R^2)\left\{1+\frac{\alpha_s(\mu_R^2)}{\pi} 
\frac{3}{2}C_F \ln\left(\frac{\mu_R^2}{m_H^2}\right)
+\frac{\alpha_s^2(\mu_R^2)}{\pi^2} \left[\left(\frac{9}{8} C_F^2
-\frac{3\beta_0}{16}C_F\right) 
\ln^2\left(\frac{\mu_R^2}{m_H^2}\right) \right. \right.
\nonumber \\ &&  \hspace{15mm} \left. \left.
{}+\left(\frac{3}{16} C_F^2+\frac{97}{48} C_F C_A-\frac{5}{24}n_f C_F\right)
\ln\left(\frac{\mu_R^2}{m_H^2}\right)\right] 
+{\cal O}\left(\alpha_s^3(\mu_R^2)\right) \right\} \, .
\eeqa

\renewcommand{\theequation}{B.\arabic{equation}}
\setcounter{equation}{0}
\section*{Appendix B: Soft and collinear 
corrections for the Drell-Yan process through NNNLO}

The complete NLO soft and virtual corrections are 
\beqa
{\hat{\sigma}}^{(1)\, SV}_{DY} &=& F^B_{DY}  
\frac{\alpha_s(\mu_R^2)}{\pi}
\left\{4 C_F\, {\cal D}_1(z) -2 C_F \ln\left(\frac{\mu_F^2}{Q^2}\right)\,  
{\cal D}_0(z) \right.
\nonumber \\ && \left.
{}+\left[2(-2+\zeta_2) C_F-\frac{3}{2} C_F  
\ln\left(\frac{\mu_F^2}{Q^2}\right)\right]\,  \delta(1-z)\right\} 
\label{NLODY}
\eeqa
where $F^B_{DY}$ is the Born term, whose exact form depends on which 
gauge boson is involved in the process. 

The leading and next-to-leading collinear corrections at NLO are 
\beq
{\hat{\sigma}}^{(1)\, C}_{DY}= F^B_{DY} \frac{\alpha_s(\mu_R^2)}{\pi}
\left\{-4 C_F\, \ln(1-z) +2 C_F \ln\left(\frac{\mu_F^2}{Q^2}\right)
+2C_F\right\} \, . 
\eeq

The complete NNLO soft and virtual corrections are
\beqa
{\hat{\sigma}}^{(2)\, SV}_{DY}&=&F^B_{DY}  
\frac{\alpha_s^2(\mu_R^2)}{\pi^2} 
\left\{8 C_F^2 \, {\cal D}_3(z)
+\left[-\frac{11}{3}C_F C_A+\frac{2}{3}C_F n_f-12 C_F^2
\ln\left(\frac{\mu_F^2}{Q^2}\right)\right] \, 
{\cal D}_2(z) \right.
\nonumber \\ && \hspace{-10mm}
{}+\left[-8 C_F^2(2+\zeta_2)+\left(\frac{67}{9}-2\zeta_2\right)C_F C_A
-\frac{10}{9} C_F n_f+4 C_F^2\ln^2\left(\frac{\mu_F^2}{Q^2}\right) \right.
\nonumber \\ && \hspace{-10mm} \quad \left.
{}-6 C_F^2 \ln\left(\frac{\mu_F^2}{Q^2}\right) 
+C_F\left(\frac{11}{3}C_A -\frac{2}{3}n_f\right)
\ln\left(\frac{\mu_R^2}{Q^2}\right)\right] \,
{\cal D}_1(z)
\nonumber \\ && \hspace{-10mm}
{}+\left[16 \zeta_3 C_F^2+C_F C_A \left(-\frac{101}{27}+\frac{11}{3}\zeta_2
+\frac{7}{2}\zeta_3\right)+\frac{2}{3}C_F n_f\left(\frac{7}{9}-\zeta_2\right)
\right.
\nonumber \\ && \hspace{-10mm} \quad \!
{}+C_F\left(3C_F+\frac{11}{12}C_A-\frac{n_f}{6}\right) 
\ln^2\left(\frac{\mu_F^2}{Q^2}\right)
-C_F\left(\frac{11}{6}C_A-\frac{n_f}{3}\right)
\ln\left(\frac{\mu_F^2}{Q^2}\right)
\ln\left(\frac{\mu_R^2}{Q^2}\right)
\nonumber \\ && \hspace{-10mm} \quad \left. 
{}+\left((8+4\zeta_2)C_F^2+(-\frac{67}{18}+\zeta_2)C_F C_A
+\frac{5}{9}C_F n_f\right)  
\ln\left(\frac{\mu_F^2}{Q^2}\right) \right] \, 
{\cal D}_0(z) 
\nonumber \\ && \hspace{-10mm}
{}+\left[C_F^2\left(\frac{511}{64}-\frac{35}{8}\zeta_2-\frac{15}{4}\zeta_3
+\frac{1}{10} \zeta_2^2\right)+C_F C_A\left(-\frac{1535}{192}
+\frac{37}{9}\zeta_2+\frac{7}{4}\zeta_3-\frac{3}{20}\zeta_2^2\right)
\right.
\nonumber \\ && \hspace{-10mm} \quad 
{}+C_F n_f \left(\frac{127}{96}-\frac{7}{9}\zeta_2+\frac{1}{2}\zeta_3\right)
+\left(C_F^2(\frac{9}{8}-2\zeta_2)+\frac{11}{16}C_F C_A
-\frac{1}{8} C_F n_f\right) \ln^2\left(\frac{\mu_F^2}{Q^2}\right) 
\nonumber \\ && \hspace{-10mm} \quad 
{}+\left(C_F^2(\frac{93}{16}-\frac{3}{2}\zeta_2-11\zeta_3)
+C_F C_A (-\frac{17}{48}-\frac{11}{6}\zeta_2+\frac{3}{2}\zeta_3)
+\frac{1}{3} C_F n_f (\frac{1}{8}+\zeta_2)\right)
\ln\left(\frac{\mu_F^2}{Q^2}\right) 
\nonumber \\ && \hspace{-10mm} \quad 
{}+\left(-\frac{11}{8}C_F C_A+\frac{1}{4} C_F n_f\right)
\ln\left(\frac{\mu_F^2}{Q^2}\right) \ln\left(\frac{\mu_R^2}{Q^2}\right)
\nonumber \\ && \hspace{-10mm} \quad \left. \left.
{}+\left(\frac{11}{6}C_F C_A (-2+\zeta_2)+\frac{1}{3} C_F n_f (2-\zeta_2)
\right)\ln\left(\frac{\mu_R^2}{Q^2}\right)
\right] \delta(1-z)\right\} \, . 
\label{NNLODY}
\eeqa
This in agreement with Refs. \cite{DY,DYSV} where the virtual terms were 
calculated in \cite{DYSV}.

The leading and some subleading collinear corrections at NNLO are
\beqa
{\hat{\sigma}}^{(2)\, C}_{DY}&=&F^B_{DY}
\frac{\alpha_s^2(\mu_R^2)}{\pi^2} 
\left\{-8 C_F^2 \, \ln^3(1-z) \right.
\nonumber \\ && \hspace{-10mm}
{}+\left[12 C_F^2+\frac{11}{3}C_F C_A-\frac{2}{3}C_F n_f+12 C_F^2
\ln\left(\frac{\mu_F^2}{Q^2}\right)\right] \, \ln^2(1-z) 
\nonumber \\ && \hspace{-10mm}
{}+\left[8C_F^2(2+\zeta_2)+C_F C_A\left(2\zeta_2-\frac{100}{9}\right)
+\frac{16}{9}n_f C_F \right.
\nonumber \\ && \hspace{-5mm} \left.
{}-4 C_F^2\ln^2\left(\frac{\mu_F^2}{Q^2}\right)  
-C_F\left(\frac{11}{3}C_A -\frac{2}{3}n_f\right)
\ln\left(\frac{\mu_R^2}{Q^2}\right)
-6 C_F^2 \ln\left(\frac{\mu_F^2}{Q^2}\right)\right] \, \ln(1-z)
\nonumber \\ && \hspace{-10mm} 
{}-C_F\left(C_F+\frac{11}{12} C_A-\frac{n_f}{6}\right)
\ln^2\left(\frac{\mu_F^2}{Q^2}\right)
+C_F\left(\frac{11}{6}C_A-\frac{n_f}{3}\right)
\ln\left(\frac{\mu_F^2}{Q^2}\right)
\ln\left(\frac{\mu_R^2}{Q^2}\right)
\nonumber \\ && \hspace{-10mm}  
{}+\left(-11 C_F^2+\frac{67}{18}C_F C_A-4\zeta_2 C_F^2-C_F C_A \zeta_2
-\frac{5}{9} n_f C_F\right)\ln\left(\frac{\mu_F^2}{Q^2}\right)
\nonumber \\ && \hspace{-10mm} \left. 
{}+C_F\left(\frac{11}{6}C_A -\frac{n_f}{3}\right)
\ln\left(\frac{\mu_R^2}{Q^2}\right)
\right\} \, . 
\eeqa
Several remarks are in order here. The leading collinear (LC) logarithms 
(i.e. $\ln^3(1-z)$) are complete. The next-to-leading collinear (NLC) 
logarithms (i.e $\ln^2(1-z)$) are not complete. 
However, numerically they are an 
excellent approximation to the complete NLC terms (84\% at $\mu_F=Q$). 
Also analytically 
the $C_F C_A$, $n_f C_F$ and the $\ln(\mu_F^2/Q^2)$ terms are exact. 
The difference of exact minus approximate NLC terms is $7 C_F^2/2$.
The next-to-next-to-leading collinear (NNLC) logarithms 
(i.e $\ln(1-z)$) are also not complete. However, again  numerically they are 
an excellent approximation to the complete NNLC terms (107\% at $\mu_F=Q$). 
Also analytically the $C_F^2 \zeta_2$, $C_F C_A \zeta_2$, $\ln^2(\mu_F^2/Q^2)$,
and $\ln(\mu_R^2/Q^2)$ terms are exact. The difference of exact minus 
approximate NNLC terms is 
$7 C_F^2/4-\beta_0 C_F+5C_F C_A/4-2C_F^2\ln(\mu_F^2/Q^2)$.
Finally, in the constant terms we only show the scale-dependent terms; the 
$\ln^2(\mu_F^2/Q^2)$, $\ln(\mu_F^2/Q^2) \ln(\mu_R^2/Q^2)$ and 
$\ln(\mu_R^2/Q^2)$ terms are exact while the  $\ln(\mu_F^2/Q^2)$ terms are 
almost exact. We note that we can even get some additional constant terms 
such as $-16 C_F^2 \zeta_3$.

The complete NNNLO soft-gluon corrections are
\beqa
{\hat{\sigma}}^{(3)\, S}_{DY}&=& 
F^B_{DY} \frac{\alpha_s^3(\mu_R^2)}{\pi^3} \;  
\left \{8 C_F^3 \; {\cal D}_5(z)
+\left[-\frac{110}{9} C_F^2 C_A+\frac{20}{9} C_F^2 n_f 
-20 C_F^3 \ln\left(\frac{\mu_F^2}{Q^2}\right) \right] \; {\cal D}_4(z) 
\right.
\nonumber \\ && \hspace{-15mm}
{}+\left[-16C_F^3 (2+3 \zeta_2) + C_F^2 C_A \left(\frac{268}{9}-8\zeta_2\right)
+\frac{121}{27} C_F C_A^2-\frac{40}{9}C_F^2 n_f
-\frac{44}{27} C_F C_A n_f+\frac{4}{27} C_F n_f^2\right.
\nonumber \\ && \hspace{-10mm}
{}+16 C_F^3 \ln^2\left(\frac{\mu_F^2}{Q^2}\right)
+\left(-12 C_F^3+\frac{88}{9} C_F^2 C_A -\frac{16}{9} C_F^2 n_f\right) 
\ln\left(\frac{\mu_F^2}{Q^2}\right) 
\nonumber \\ && \hspace{-10mm}\left.
{}+\left(\frac{44}{3} C_F^2 C_A -\frac{8}{3} C_F^2 n_f\right) 
\ln\left(\frac{\mu_R^2}{Q^2}\right)\right] \; {\cal D}_3(z) 
\nonumber \\ && \hspace{-15mm}
{}+\left[160 \zeta_3 C_F^3+C_F^2 C_A\left(-\frac{70}{9}+\frac{176}{3}\zeta_2
+21 \zeta_3\right)+C_F C_A^2\left(-\frac{445}{27}+\frac{11}{3}\zeta_2\right)
+C_F^2 n_f\left(\frac{17}{18}-\frac{32}{3}\zeta_2\right)
\right.
\nonumber \\ && \hspace{-10mm}
{}+C_F C_A n_f\left(\frac{289}{54}-\frac{2}{3}\zeta_2\right)
-\frac{10}{27}C_F n_f^2-4C_F^3\ln^3\left(\frac{\mu_F^2}{Q^2}\right)
+\left(18 C_F^3+\frac{11}{2} C_F^2 C_A-C_F^2 n_f\right) 
\ln^2\left(\frac{\mu_F^2}{Q^2}\right)
\nonumber \\ && \hspace{-10mm}
{}+\left(24 (2+3\zeta_2)C_F^3+(-\frac{235}{6}+12\zeta_2)C_F^2 C_A
+\frac{17}{3}C_F^2 n_f\right) \ln\left(\frac{\mu_F^2}{Q^2}\right)
\nonumber \\ && \hspace{-10mm} 
{}+(-22 C_F^2 C_A+4C_F^2 n_f) \ln\left(\frac{\mu_F^2}{Q^2}\right)
\ln\left(\frac{\mu_R^2}{Q^2}\right)
\nonumber \\ && \hspace{-10mm} \left.
{}+\left(-\frac{121}{18} C_F C_A^2
+\frac{22}{9} C_F C_A n_f-\frac{2}{9} C_F n_f^2\right)
\ln\left(\frac{\mu_R^2}{Q^2}\right)\right] \; {\cal D}_2(z) 
\nonumber \\ && \hspace{-15mm}
{}+\left[C_F^3\left(\frac{511}{16}+\frac{93}{2}\zeta_2-15\zeta_3
-192\zeta_4+\frac{162}{5}\zeta_2^2\right)
+C_F^2 C_A\left(-\frac{8893}{144}-\frac{182}{9}\zeta_2-81\zeta_3
+\frac{57}{5}\zeta_2^2\right)\right.
\nonumber \\ && \hspace{-10mm}
{}+C_F C_A^2\left(\frac{15503}{648}-\frac{188}{9}\zeta_2-11\zeta_3
+\frac{11}{5}\zeta_2^2\right)
+C_F^2 n_f\left(\frac{67}{9}+\frac{32}{9}\zeta_2+20\zeta_3\right)
\nonumber \\ && \hspace{-10mm}
{}+C_F C_A n_f\left(-\frac{2051}{324}+6\zeta_2\right)
+C_F n_f^2\left(\frac{25}{81}-\frac{4}{9}\zeta_2\right)
+\left(-6C_F^3-\frac{11}{3}C_F^2 C_A+\frac{2}{3} C_F^2 n_f\right)
\ln^3\left(\frac{\mu_F^2}{Q^2}\right)
\nonumber \\ && \hspace{-10mm}
{}+\left(-(\frac{23}{2}+32\zeta_2)C_F^3+(\frac{635}{36}-4\zeta_2)C_F^2 C_A
-\frac{49}{18} C_F^2 n_f\right)\ln^2\left(\frac{\mu_F^2}{Q^2}\right)
\nonumber \\ && \hspace{-10mm}
{}+\frac{2}{3}C_F^2(11C_A-2n_f)\ln^2\left(\frac{\mu_F^2}{Q^2}\right)
\ln\left(\frac{\mu_R^2}{Q^2}\right)
+C_F^2(-11C_A+2n_f)\ln\left(\frac{\mu_F^2}{Q^2}\right)
\ln\left(\frac{\mu_R^2}{Q^2}\right)
\nonumber \\ && \hspace{-10mm}
{}+\left((\frac{93}{4}+18\zeta_2-172 \zeta_3) C_F^3
+(\frac{257}{108}-\frac{101}{3}\zeta_2-8\zeta_3)C_F^2 C_A
+(-\frac{13}{54}+\frac{20}{3}\zeta_2)C_F^2 n_f\right)
\ln\left(\frac{\mu_F^2}{Q^2}\right)
\nonumber \\ && \hspace{-10mm}
{}+\left(\frac{121}{36}C_F C_A^2-\frac{11}{9} C_F C_A n_f
+\frac{1}{9} C_F n_f^2\right)\ln^2\left(\frac{\mu_R^2}{Q^2}\right)
\nonumber \\ && \hspace{-10mm}
{}+\left(-\frac{44}{3} (2+\zeta_2)C_F^2 C_A
+(\frac{445}{27}-\frac{11}{3}\zeta_2)C_F C_A^2
+\frac{1}{6}(29+16\zeta_2) C_F^2 n_f \right.
\nonumber \\ && \left. \left. 
+(-\frac{289}{54}+\frac{2}{3}\zeta_2)C_F C_A n_f 
+\frac{10}{27} C_F n_f^2\right)\ln\left(\frac{\mu_R^2}{Q^2}\right)\right] 
\; {\cal D}_1(z) 
\nonumber \\ && \hspace{-15mm}
{}+\left[C_F^3(-64\zeta_3+192 \zeta_5-96 \zeta_2 \zeta_3)
+C_F^2 C_A\left(\frac{404}{27}-\frac{194}{27}\zeta_2+\frac{410}{9}\zeta_3
+\frac{220}{3}\zeta_4-22\zeta_2^2-23\zeta_2\zeta_3\right) \right.
\nonumber \\ && \hspace{-10mm}
{}+C_F C_A^2 \left(-\frac{297029}{23328}+\frac{6139}{324}\zeta_2
+\frac{2509}{108}\zeta_3
-6\zeta_5-\frac{187}{60}\zeta_2^2-\frac{11}{6}\zeta_2\zeta_3\right) 
\nonumber \\ && \hspace{-10mm}
{}+C_F^2 n_f \left(-\frac{3}{32}+\frac{61}{54}\zeta_2-\frac{179}{18}\zeta_3
-\frac{40}{3} \zeta_4 +\frac{19}{5}\zeta_2^2\right) 
+C_F C_A n_f \left(\frac{31313}{11664}-\frac{1837}{324}\zeta_2
-\frac{155}{36}\zeta_3+\frac{23}{30}\zeta_2^2\right) 
\nonumber \\ && \hspace{-10mm}
{}+C_F n_f^2 \left(-\frac{58}{729}+\frac{10}{27}\zeta_2
+\frac{5}{27}\zeta_3\right) 
\nonumber \\ && \hspace{-10mm}
{}+\left((-\frac{9}{4}+4\zeta_2) C_F^3-\frac{11}{4}C_F^2 C_A
-\frac{121}{216}C_F C_A^2+\frac{1}{2}C_F^2 n_f+\frac{11}{54} C_F C_A n_f
-\frac{1}{54}C_F n_f^2\right)\ln^3\left(\frac{\mu_F^2}{Q^2}\right)
\nonumber \\ && \hspace{-10mm}
{}+\left((-\frac{93}{8}-9\zeta_2+38 \zeta_3) C_F^3
+(\frac{21}{8}+\frac{\zeta_2}{3}-3\zeta_3)C_F^2 C_A
+(\frac{445}{108}-\frac{11}{12}\zeta_2)C_F C_A^2 \right. 
\nonumber \\ && \hspace{-10mm} \left.
{}+(-\frac{3}{8}-\frac{\zeta_2}{3}) C_F^2 n_f
+(-\frac{289}{216}+\frac{\zeta_2}{6}) C_F C_A n_f
+\frac{5}{54} C_F n_f^2\right)
\ln^2\left(\frac{\mu_F^2}{Q^2}\right)
\nonumber \\ && \hspace{-10mm}
{}+\left(\frac{11}{2} C_F^2 C_A +\frac{121}{72}C_F C_A^2 -C_F^2 n_f
-\frac{11}{18} C_F C_A n_f+\frac{1}{18} C_F n_f^2\right)
\ln^2\left(\frac{\mu_F^2}{Q^2}\right)\ln\left(\frac{\mu_R^2}{Q^2}\right)
\nonumber \\ && \hspace{-10mm}
{}+\left(-\frac{121}{72} C_F C_A^2+\frac{11}{18} C_F C_A n_f
-\frac{1}{18} C_F n_f^2 \right)
\ln\left(\frac{\mu_F^2}{Q^2}\right)\ln^2\left(\frac{\mu_R^2}{Q^2}\right)
\nonumber \\ && \hspace{-10mm}
{}+\left((\frac{44}{3}+\frac{22}{3}\zeta_2)C_F^2 C_A
+(-\frac{445}{54}+\frac{11}{6}\zeta_2) C_F C_A^2
+(-\frac{29}{12}-\frac{4}{3}\zeta_2)C_F^2 n_f \right.
\nonumber \\ && \left. 
+(\frac{289}{108}-\frac{\zeta_2}{3}) C_F C_A n_f
-\frac{5}{27} C_F n_f^2 \right)
\ln\left(\frac{\mu_F^2}{Q^2}\right)\ln\left(\frac{\mu_R^2}{Q^2}\right)
\nonumber \\ && \hspace{-10mm}
{}+\left((-\frac{511}{32}-\frac{93}{4}\zeta_2-\frac{33}{2} \zeta_3
+96\zeta_4-\frac{81}{5}\zeta_2^2) C_F^3 
+(\frac{3503}{96}+\frac{83}{18}\zeta_2+\frac{71}{12}\zeta_3
-\frac{57}{10}\zeta_2^2) C_F^2 C_A \right.
\nonumber \\ &&
{}+(-\frac{245}{48}+\frac{67}{18}\zeta_2-\frac{11}{12}\zeta_3
-\frac{11}{10}\zeta_2^2)C_F C_A^2
+(-\frac{9}{2}-\frac{7}{9}\zeta_2-\frac{14}{3}\zeta_3) C_F^2 n_f
\nonumber \\ && \left.
{}+(\frac{209}{216}-\frac{5}{9}\zeta_2+\frac{7}{6}\zeta_3)C_F C_A n_f
+\frac{1}{54} C_F n_f^2 \right) \ln\left(\frac{\mu_F^2}{Q^2}\right)
\nonumber \\ && \hspace{-10mm}
{}+\left(\frac{88}{3} \zeta_3 C_F^2 C_A 
+(-\frac{1111}{162}+\frac{121}{18}\zeta_2+\frac{77}{12}\zeta_3) C_F C_A^2
-\frac{16}{3}\zeta_3 C_F^2 n_f \right.
\nonumber \\ && \hspace{-10mm}\left. \left. \left.
{}+(\frac{178}{81}-\frac{22}{9}\zeta_2-\frac{7}{6} \zeta_3) C_F C_A n_f 
+(-\frac{14}{81}+\frac{2}{9}\zeta_2) C_F n_f^2
\right) \ln\left(\frac{\mu_R^2}{Q^2}\right)\right]\; {\cal D}_0(z) \right\}\, .
\label{NNNLODY}
\eeqa
The scale-independent terms in the above expression are in agreement 
with Refs. \cite{VR,MVNNNLO}.

The leading and some subleading collinear corrections at NNNLO are
\beqa
{\hat{\sigma}}^{(3)\, C}_{DY}&=& 
F^B_{DY} \frac{\alpha_s^3(\mu_R^2)}{\pi^3} \;  
\left \{-8 C_F^3 \, \ln^5(1-z)
+\left[\frac{110}{9} C_F^2 C_A-\frac{20}{9} C_F^2 n_f 
+20 C_F^3 \ln\left(\frac{\mu_F^2}{Q^2}\right) \right] \, \ln^4(1-z) 
\right.
\nonumber \\ && \hspace{-15mm} 
{}+\left[16 C_F^3 (2+3\zeta_2)
+C_F^2 C_A\left(-\frac{488}{9}+8\zeta_2\right)
-\frac{121}{27} C_F C_A^2+C_F\frac{n_f}{27}(240C_F+44 C_A-4n_f) \right. 
\nonumber \\ && \hspace{-10mm} 
{}-16 C_F^3 \ln^2\left(\frac{\mu_F^2}{Q^2}\right)
+C_F^2\left(12 C_F-\frac{88}{9} C_A+\frac{16}{9}n_f\right)
\ln\left(\frac{\mu_F^2}{m_H^2}\right)
\nonumber \\ && \hspace{-10mm} \left.
{}+\left(-\frac{44}{3} C_F^2 C_A +\frac{8}{3} C_F^2 n_f\right) 
\ln\left(\frac{\mu_R^2}{Q^2}\right)\right] \, \ln^3(1-z) 
\nonumber \\ && \hspace{-15mm} \left.
{}+\left[4 C_F^3 \ln^3\left(\frac{\mu_F^2}{Q^2}\right)
+C_F^2 (22 C_A -4 n_f) \ln\left(\frac{\mu_F^2}{Q^2}\right)
\ln\left(\frac{\mu_R^2}{Q^2}\right)\right] \, \ln^2(1-z) \right\} \, . 
\eeqa
Again we note that only the LC terms (i.e. $\ln^5(1-z)$) are complete. 
The NLC (i.e. $\ln^4(1-z)$) and NNLC (i.e. $\ln^3(1-z)$) terms are not 
complete but, based on our study at NNLO, 
we expect them to be a very good approximation to the complete terms.
In the $\ln^2(1-z)$ terms we only show the scale terms that we expect to be 
exact at this accuracy.

\renewcommand{\theequation}{C.\arabic{equation}}
\setcounter{equation}{0}
\section*{Appendix C: Soft and collinear corrections 
for $gg \rightarrow H$ through NNNLO}

The NLO soft and virtual corrections are 
\beqa
{\hat{\sigma}}^{(1)\, SV}_{gg \rightarrow H} &=& 
F^B_{gg \rightarrow H}(\mu_R^2) \frac{\alpha_s(\mu_R^2)}{\pi}
\left\{4C_A \, {\cal D}_1(z)  
-2 C_A \ln\left(\frac{\mu_F^2}{m_H^2}\right)\, {\cal D}_0(z) \right.
\nonumber \\ && \left.
{}+\left[\frac{5}{2}C_A-\frac{3}{2}C_F+2 C_A \zeta_2 
+\left(\frac{11}{6} C_A -\frac{1}{3} n_f\right) 
\ln\left(\frac{\mu_R^2}{\mu_F^2}\right) \right]\, \delta(1-z)\right\} 
\label{NLOggH}
\eeqa
with 
\beq
F^B_{gg \rightarrow H}(\mu_R^2)=\frac{\alpha_s^2(\mu_R^2)}{\pi^2}
\frac{\pi}{72 (N_c^2-1) v^2}
\eeq
where $v=246$ GeV is the Higgs boson vacuum expectation value.

The leading and next-to-leading collinear corrections at NLO are 
\beq
{\hat{\sigma}}^{(1)\, C}_{gg \rightarrow H}= 
F^B_{gg \rightarrow H}(\mu_R^2) \frac{\alpha_s(\mu_R^2)}{\pi}
\left\{-8 C_A\, \ln(1-z) +4 C_A \ln\left(\frac{\mu_F^2}{m_H^2}\right)
+2 C_A\right\}
\, . 
\eeq

The complete NNLO soft and virtual corrections are
\beqa
{\hat{\sigma}}^{(2)\, SV}_{gg \rightarrow H}&=&F^B_{gg 
\rightarrow H}(\mu_R^2) \frac{\alpha_s^2(\mu_R^2)}{\pi^2} 
\left\{8 C_A^2 \, {\cal D}_3(z)
+\left[-\frac{11}{3}C_A^2+\frac{2}{3}C_A n_f-12 C_A^2
\ln\left(\frac{\mu_F^2}{m_H^2}\right)\right] \, 
{\cal D}_2(z) \right.
\nonumber \\ && 
{}+\left[\left(\frac{157}{9}-10\zeta_2\right) C_A^2-6 C_F C_A
-\frac{10}{9} C_A n_f+4 C_A^2\ln^2\left(\frac{\mu_F^2}{m_H^2}\right) \right.
\nonumber \\ && \quad \left.
{}-C_A\left(\frac{22}{3}C_A -\frac{4}{3}n_f\right) 
\ln\left(\frac{\mu_F^2}{m_H^2}\right) 
+C_A\left(11C_A - 2n_f\right)
\ln\left(\frac{\mu_R^2}{m_H^2}\right)\right] \,
{\cal D}_1(z)
\nonumber \\ && 
{}+\left[C_A^2\left(-\frac{101}{27}+\frac{11}{3}\zeta_2
+\frac{39}{2}\zeta_3\right)+C_A n_f\left(\frac{14}{27}
-\frac{2}{3}\zeta_2\right)
\right.
\nonumber \\ && \quad  
{}+C_A\left(\frac{55}{12} C_A-\frac{5}{6}n_f\right) 
\ln^2\left(\frac{\mu_F^2}{m_H^2}\right)
-C_A\left(\frac{11}{2}C_A-n_f\right)
\ln\left(\frac{\mu_F^2}{m_H^2}\right)
\ln\left(\frac{\mu_R^2}{m_H^2}\right)
\nonumber \\ && \quad \left. 
{}+\left(-\frac{157}{18}C_A^2+5\zeta_2 C_A^2+3C_F C_A
+\frac{5}{9}C_A n_f\right)\ln\left(\frac{\mu_F^2}{m_H^2}\right) \right] \, 
{\cal D}_0(z) 
\nonumber \\ &&  
{}+\left[C_A^2\left(\frac{3187}{288}+\frac{157}{18}\zeta_2-\frac{55}{12}\zeta_3
-\frac{\zeta_2^2}{20}\right)+\frac{9}{4}C_F^2-C_F C_A\left(\frac{145}{24}
+3\zeta_2\right)-\frac{C_F}{12}-\frac{5}{96}C_A \right.
\nonumber \\ && \quad 
{}-C_A n_f \left(\frac{287}{144}+\frac{5}{9}\zeta_2
+\frac{\zeta_3}{6}\right)-C_F n_f \left(\frac{41}{24}-\zeta_3\right) 
+\left(\frac{7}{8}C_A^2-\frac{11}{8}C_F C_A+\frac{n_f}{2}C_F\right)
\ln\left(\frac{m_H^2}{m_t^2}\right)
\nonumber \\ && \quad 
{}+\left(\frac{121}{48} C_A^2-\frac{11}{12}C_A n_f+\frac{n_f^2}{12}
\right)\ln^2\left(\frac{\mu_F^2}{\mu_R^2}\right)
-2 \zeta_2 C_A^2 \ln^2\left(\frac{\mu_F^2}{m_H^2}\right)
\nonumber \\ && \quad 
{}+\left(-(\frac{71}{12}+\frac{11}{3}\zeta_2+\frac{19}{2}\zeta_3)C_A^2 
+\frac{11}{4} C_F C_A 
+(\frac{7}{6}+\frac{2}{3}\zeta_2)C_A n_f-\frac{1}{4}C_F n_f \right)
\ln\left(\frac{\mu_F^2}{m_H^2}\right)
\nonumber \\ && \quad \! \! \left.\left.
{}+\left((\frac{199}{24}+\frac{11}{2}\zeta_2)C_A^2-\frac{33}{8} C_F C_A 
-(\frac{5}{3}+\zeta_2)C_A n_f+\frac{1}{2}C_F n_f \right)
\ln\left(\frac{\mu_R^2}{m_H^2}\right)\right]
\delta(1-z)\right\} \, ,
\label{NNLOggH}
\eeqa
where $m_t$ is the top quark mass.
This is in agreement with \cite{HK1,CAKM,RSvN,CFG}, and the virtual
corrections were calculated in \cite{H2l}.

The leading and some subleading collinear corrections at NNLO are
\beqa
{\hat{\sigma}}^{(2)\, C}_{gg \rightarrow H}&=&
F^B_{gg \rightarrow H}(\mu_R^2) \frac{\alpha_s^2(\mu_R^2)}{\pi^2} 
\left\{-16 C_A^2 \, \ln^3(1-z) 
+\left[\frac{58}{3}C_A^2-\frac{4}{3}C_A n_f+24 C_A^2
\ln\left(\frac{\mu_F^2}{m_H^2}\right)\right] \, \ln^2(1-z) \right.
\nonumber \\ && \hspace{-10mm} \left.
{}+{\cal O} (\ln(1-z)) \right\} \, .
\eeqa
Again, only the leading collinear (LC) logarithms 
(i.e. $\ln^3(1-z)$) are complete. The next-to-leading collinear (NLC) 
logarithms (i.e $\ln^2(1-z)$) are not complete. 
However, numerically they are an 
excellent approximation to the complete NLC terms (83\% at $\mu_F=m_H$). 
Also analytically the $n_f C_A$ and the $\ln(\mu_F^2/m_H^2)$ terms are exact. 
The difference of exact minus approximate NLC terms is $7 C_A^2/2$.

The NNNLO soft-gluon corrections are
\beqa
{\hat{\sigma}}^{(3)\, S}_{gg \rightarrow H}&=& 
F^B_{gg \rightarrow H}(\mu_R^2) \frac{\alpha_s^3(\mu_R^2)}{\pi^3} \;  
\left \{8 C_A^3 \; {\cal D}_5(z)
+\left[-\frac{110}{9} C_A^3+\frac{20}{9} C_A^2 n_f 
-20 C_A^3 \ln\left(\frac{\mu_F^2}{m_H^2}\right) \right] \; {\cal D}_4(z) 
\right.
\nonumber \\ && \hspace{-15mm}
{}+\left[C_A^3 \left(\frac{1465}{27}-56\zeta_2\right)-12 C_A^2 C_F
-\frac{164}{27}C_A^2 n_f+\frac{4}{27} C_A n_f^2
+16 C_A^3 \ln^2\left(\frac{\mu_F^2}{m_H^2}\right) \right.
\nonumber \\ && \hspace{-10mm} \left.
{}+\left(-\frac{44}{9} C_A^3 +\frac{8}{9} C_A^2 n_f\right) 
\ln\left(\frac{\mu_F^2}{m_H^2}\right) 
+\left(\frac{88}{3} C_A^3 -\frac{16}{3} C_A^2 n_f\right) 
\ln\left(\frac{\mu_R^2}{m_H^2}\right)\right] \; {\cal D}_3(z) 
\nonumber \\ && \hspace{-15mm}
{}+\left[C_A^3\left(-\frac{2597}{54}+\frac{187}{3}\zeta_2+181 \zeta_3\right)
+\frac{11}{2} C_A^2 C_F
+C_A^2  n_f\left(\frac{547}{54}-\frac{34}{3}\zeta_2\right)
\right.
\nonumber \\ && \hspace{-10mm}
{}-\frac{1}{2} C_F C_A n_f
-\frac{10}{27}C_A n_f^2-4 C_A^3 \ln^3\left(\frac{\mu_F^2}{m_H^2}\right)
+\left(\frac{55}{2} C_A^3 - 5 C_A^2 n_f\right) 
\ln^2\left(\frac{\mu_F^2}{m_H^2}\right)
\nonumber \\ && \hspace{-10mm}
{}+\left((-\frac{1223}{18}+84\zeta_2)C_A^3+18 C_A^2 C_F
+\frac{38}{9}C_A^2 n_f+\frac{2}{9} C_A n_f^2\right) 
\ln\left(\frac{\mu_F^2}{m_H^2}\right)
\nonumber \\ && \hspace{-10mm} \left.
{}+(-44 C_A^3+8 C_A^2 n_f) \ln\left(\frac{\mu_F^2}{m_H^2}\right)
\ln\left(\frac{\mu_R^2}{m_H^2}\right)
+\left(-\frac{121}{9} C_A^3
+\frac{44}{9} C_A^2 n_f-\frac{4}{9} C_A n_f^2\right)
\ln\left(\frac{\mu_R^2}{m_H^2}\right)\right] {\cal D}_2(z) 
\nonumber \\ && \hspace{-15mm}
{}+\left[C_A^3\left(\frac{28123}{324}
-\frac{227}{3}\zeta_2-\frac{352}{3}\zeta_3-192\zeta_4+46\zeta_2^2\right)
+C_A^2 C_F \left(-\frac{106}{3}+15\zeta_2\right)+9 C_A C_F^2 \right.
\nonumber \\ && \hspace{-10mm}
{}-\frac{5}{24}C_A^2-\frac{1}{3}C_F C_A+\left(\frac{7}{2}C_A^3
-\frac{11}{2} C_A^2 C_F+2 C_A C_F n_f\right)
\ln\left(\frac{m_H^2}{m_t^2}\right)
\nonumber \\ && \hspace{-10mm}
{}+C_A^2 n_f\left(-\frac{2767}{162}+\frac{94}{9}\zeta_2
+\frac{46}{3}\zeta_3\right)
+C_F C_A n_f\left(-\frac{179}{24}+6 \zeta_3\right)
+C_A n_f^2\left(\frac{25}{81} -\frac{4}{9}\zeta_2\right)
\nonumber \\ && \hspace{-10mm}
{}+\left(-11 C_A^3+2 C_A^2 n_f\right)
\ln^3\left(\frac{\mu_F^2}{m_H^2}\right)
+\left((\frac{1259}{36}-36\zeta_2)C_A^3-6 C_A^2 C_F
-\frac{53}{9} C_A^2 n_f+\frac{1}{3} C_A n_f^2\right)
\ln^2\left(\frac{\mu_F^2}{m_H^2}\right)
\nonumber \\ && \hspace{-10mm}
{}+\frac{4}{3}C_A^2(11C_A-2n_f)\ln^2\left(\frac{\mu_F^2}{m_H^2}\right)
\ln\left(\frac{\mu_R^2}{m_H^2}\right)
+\left(-\frac{242}{9} C_A^3-\frac{8}{9} C_A n_f^2
+\frac{88}{9} C_A^2 n_f\right)
\ln\left(\frac{\mu_F^2}{m_H^2}\right)
\ln\left(\frac{\mu_R^2}{m_H^2}\right)
\nonumber \\ && \hspace{-10mm}
{}+\left((-\frac{1207}{54}-11\zeta_2-180 \zeta_3) C_A^3
+11 C_A^2 C_F+(\frac{64}{9}+2 \zeta_2)C_A^2 n_f -C_F C_A n_f
-\frac{10}{27} C_A n_f^2\right)
\ln\left(\frac{\mu_F^2}{m_H^2}\right)
\nonumber \\ && \hspace{-10mm}
{}+\left(\frac{121}{6} C_A^3-\frac{22}{3} C_A^2 n_f
+\frac{2}{3} C_A n_f^2\right)\ln^2\left(\frac{\mu_R^2}{m_H^2}\right)
+\left(\frac{3913}{54} C_A^3 -22 C_A^2 C_F
-\frac{110}{3}\zeta_2 C_A^3 \right.
\nonumber \\ && \hspace{-5mm} \left. \left. 
{}-\frac{983}{54} C_A^2 n_f+\frac{5}{2}C_F C_A n_f
+\frac{20}{3}\zeta_2 C_A^2 n_f 
+\frac{20}{27} C_A n_f^2\right)\ln\left(\frac{\mu_R^2}{m_H^2}\right)\right] 
\; {\cal D}_1(z) 
\nonumber \\ && \hspace{-15mm}
{}+\left[C_A^3(-\frac{515189}{23328}+\frac{11533}{324} \zeta_2
+\frac{7103}{54}\zeta_3+\frac{220}{3}\zeta_4+186 \zeta_5
-\frac{1507}{60} \zeta_2^2
-\frac{725}{6} \zeta_2 \zeta_3) \right.
\nonumber \\ && \hspace{-10mm}
{}+C_A^2 C_F \left(\frac{101}{18}-\frac{11}{2}\zeta_2-\frac{117}{4}\zeta_3
\right)+C_A^2 n_f \left(\frac{46433}{11664}-\frac{2713}{324}\zeta_2
-\frac{475}{36}\zeta_3-\frac{40}{3} \zeta_4 +\frac{143}{30}\zeta_2^2\right) 
\nonumber \\ && \hspace{-10mm}
{}+C_F C_A n_f \left(\frac{1039}{864}+\frac{\zeta_2}{2}
-\frac{19}{18}\zeta_3-\frac{1}{5}\zeta_2^2\right) 
+C_A n_f^2 \left(-\frac{58}{729}+\frac{10}{27}\zeta_2
+\frac{5}{27}\zeta_3\right) 
\nonumber \\ && \hspace{-10mm}
{}+\left((-\frac{1573}{216}+4\zeta_2) C_A^3
+\frac{143}{54}C_A^2 n_f
-\frac{13}{54}C_A n_f^2\right)\ln^3\left(\frac{\mu_F^2}{m_H^2}\right)
\nonumber \\ && \hspace{-10mm}
{}+\left((\frac{1805}{72}-\frac{143}{12}\zeta_2
+35 \zeta_3) C_A^3 -\frac{55}{8}C_A^2 C_F+\frac{5}{8} C_F C_A n_f
+(-\frac{457}{72}+\frac{13}{6} \zeta_2) C_A^2 n_f \right.
\nonumber \\ && \hspace{-5mm} \left.
{}+\frac{5}{18} C_A n_f^2\right)
\ln^2\left(\frac{\mu_F^2}{m_H^2}\right)
+\left(\frac{605}{36} C_A^3 -\frac{55}{9}C_A^2 n_f
+\frac{5}{9} C_A n_f^2\right)
\ln^2\left(\frac{\mu_F^2}{m_H^2}\right)\ln\left(\frac{\mu_R^2}{m_H^2}\right)
\nonumber \\ && \hspace{-10mm}
{}+\left(-\frac{121}{12} C_A^3+\frac{11}{3} C_A^2 n_f
-\frac{1}{3} C_A n_f^2 \right)
\ln\left(\frac{\mu_F^2}{m_H^2}\right)\ln^2\left(\frac{\mu_R^2}{m_H^2}\right)
+\left((-\frac{3913}{108}+\frac{55}{3}\zeta_2) C_A^3 \right.
\nonumber \\ && \hspace{-5mm} \left.
{}+11 C_A^2 C_F
+(\frac{983}{108}-\frac{10}{3} \zeta_2) C_A^2 n_f-\frac{5}{4} C_F C_A n_f
-\frac{10}{27} C_A n_f^2 \right)
\ln\left(\frac{\mu_F^2}{m_H^2}\right)\ln\left(\frac{\mu_R^2}{m_H^2}\right)
\nonumber \\ && \hspace{-10mm}
{}+\left((-\frac{19235}{648}+\frac{439}{18}\zeta_2-\frac{77}{6}\zeta_3
+96\zeta_4-23\zeta_2^2) C_A^3+C_A^2 C_F(\frac{53}{3}-\frac{15}{2}\zeta_2)
-\frac{9}{2}C_A C_F^2 
\right.
\nonumber \\ && \hspace{-5mm}
{}+\frac{5}{48}C_A^2+\frac{1}{6} C_A C_F
+(\frac{179}{48}-3\zeta_3) C_F C_A n_f 
+(\frac{1343}{324}-\frac{1}{3}\zeta_2+\frac{16}{3}\zeta_3) C_A^2 n_f
+(\frac{31}{162}-\frac{2}{9} \zeta_2) C_A n_f^2 
\nonumber \\ && \hspace{-5mm} \left.
{}+(-\frac{7}{4} C_A^3+\frac{11}{4} C_A^2 C_F - C_A C_F n_f)
\ln(\frac{m_H^2}{m_t^2})\right)
\ln\left(\frac{\mu_F^2}{m_H^2}\right)
\nonumber \\ && \hspace{-10mm}
{}+\left((-\frac{1111}{81}+\frac{121}{9} \zeta_2+\frac{143}{2}\zeta_3)C_A^3 
+(\frac{356}{81}-\frac{44}{9}\zeta_2-13 \zeta_3) C_A^2 n_f 
\right.
\nonumber \\ && \hspace{-5mm} \left. \left. \left.
{}+(-\frac{28}{81}+\frac{4}{9}\zeta_2) C_A n_f^2 \right) 
\ln\left(\frac{\mu_R^2}{m_H^2}\right)\right]\; {\cal D}_0(z) \right\}\, .
\label{NNNLOggH}
\eeqa
The scale-independent terms in the above expression are in agreement 
with Refs. \cite{VR,MVNNNLO}.

The leading and some subleading collinear corrections at NNNLO are
\beqa
{\hat{\sigma}}^{(3)\, C}_{gg \rightarrow H}&=& 
F^B_{gg \rightarrow H}(\mu_R^2) \frac{\alpha_s^3(\mu_R^2)}{\pi^3} \;  
\left \{-16 C_A^3 \, \ln^5(1-z)
+\left[\frac{220}{9} C_A^3-\frac{40}{9} C_A^2 n_f 
+40 C_A^3 \ln\left(\frac{\mu_F^2}{m_H^2}\right) \right] \, \ln^4(1-z) 
\right.
\nonumber \\ && \hspace{-15mm} \left.  
{}+{\cal O} \, (\ln^3(1-z)) \right\} \, . 
\eeqa
Again, only the leading terms in the above expression are exact.

\end{document}